\documentclass[aps,prd,twocolumn,preprintnumbers,nofootinbib,superscriptaddress,amsmath]{revtex4-2}
\usepackage[english]{babel}
\usepackage{bm}
\usepackage[utf8]{inputenc}
\usepackage[colorinlistoftodos, color=green!40, prependcaption]{todonotes}
\usepackage{amssymb}
\usepackage[pdftex, pdftitle={Article}, pdfauthor={Author}]{hyperref} 
\def\aap{{Astronomy and Astrophysics}}

\def\araa{{Annual Review of Astron and Astrophys}} 

\def\physscr{{Physica Scripta}}

\usepackage{graphicx}
\usepackage{dcolumn}
\usepackage{bm}
\usepackage{hyperref}
\usepackage{listings}

\begin{document}


\title{Reconstructing Cosmic History with Machine Learning: A Study Using CART, MLPR, and SVR}
\author{Agripino Sousa-Neto}
\email{agripinoneto@on.br}
\affiliation{Observatório Nacional, Rio de Janeiro - RJ, 20921-400, Brazil}

\author{Maria Aldinez Dantas}
\affiliation{Departamento de Física, Universidade do Estado do Rio Grande do Norte, 59610-210, Mossoró- RN, Brazil }%

\date{\today}

\begin{abstract}

In this work, we reconstruct cosmic history via supervised learning through three methods: Classification and Regression Trees (CART), Multi-layer Perceptron Regressor (MLPR), and Support Vector Regression (SVR). For this purpose, we use ages of simulated galaxies based on 32 massive, early-time, passively evolving galaxies in the range $0.12 < z < 1.85$, with absolute ages determined. Using this sample, we simulate subsamples of 100, 1000, 2000, 3334, 6680 points, through the Monte Carlo Method and adopting a Gaussian distribution centering on a spatially flat $\Lambda$CDM as a fiducial model. We found that the SVR method demonstrates the best performance during the process. The methods MLPR and CART also present satisfactory performance, but their mean square errors are greater than those found for the  SVR. Using the reconstructed ages, we estimate the matter density parameter and equation of state (EoS) and our analysis found the SVR with 600 predict points obtains $\Omega_m=0.329\pm{}^{0.010}_{0.010}$ and the dark energy EoS parameter $\omega= -1.054\pm{}^{0.087}_{0.126}$, which are consistent with the values from the literature. We highlight that we found the most consistent results for the subsample with 2000 points, which returns 600 predicted points and has the best performance, considering its small sample size and high accuracy. We present the reconstructed curves of galaxy ages and the best fits cosmological parameters.
\end{abstract}

\maketitle


\section{Introduction}\label{Introduction}

In the early 1990s, a key issue in cosmology was the so-called ``age tensions": estimates of the Hubble constant around $H_0 \sim 80$ km s$^{-1} $Mpc$^{-1}$ implied a Universe with an age of only about 8–10 Gyr \citep{ostriker1995cosmicconcordance}. This conflicted with the ages of the oldest known stellar populations, such as globular clusters, which were estimated to be older than 12 Gyr \citep{jimenez1997ageuniverse, Chaboyer_1998, Krauss_1995,2003Sci...299...65K}. This tension challenged the standard cosmological model and sparked intense debate about the correct values of fundamental parameters. One solution was the introduction of a cosmological constant, $\Lambda$, which helped reconcile the high Hubble constant with an older universe, leading to the success of the now-standard $\Lambda$CDM model \citep{Krauss_1995}.

This historical puzzle highlighted the critical role of accurately determining the ages of cosmic objects in constraining cosmological models. In particular, galaxies can serve as cosmic chronometers \citep{Jimenez_2002}  whose age estimates are largely model-independent, offering an alternative approach to traditional cosmological probes. By comparing galaxy ages across different redshifts \citep{simon2005constraints,nolan2001sun}, one can track the expansion history of the Universe and place constraints on key cosmological parameters, such as the Hubble constant \citep{Vagnozzi_2022,Wei_2022} or the dark energy equation of state \citep{dantas2007age,dantas2011time,dantas2009}.

However, interpreting galaxy ages in a cosmological context requires careful consideration of the time elapsed between the Big Bang and the formation of each object—a period known as the incubation time or delay factor \cite{capozziello2004constraining,pires2006lookback}. A helpful analogy is the way we measure human age: when someone asks how old you are, the answer typically refers to the time since your birth, not since conception. The time between conception (the true beginning) and birth is analogous to the delay factor in cosmology. Similarly, galaxies form some time after the Big Bang, and this delay must be taken into account when using their ages to infer cosmological parameters. 

Since different objects exhibit different delay factors \citep{binici2024age,Gao_2024,2012A&A...537A..31C}, imposing cosmological constraints directly from observed ages can be challenging. Moreover, the observational sample of galaxy ages remains limited. To address this limitation, we use simulated galaxy ages to increase the number of data points. In our simulations, we adopt the $\Lambda$CDM model as the fiducial cosmology for computing theoretical ages, which do not require incorporating the delay factor.

Testing cosmological theories demands robust statistical analyses. Given that some datasets contain relatively few data points, we require models capable of extracting meaningful interpretations. In this context, regression and reconstruction techniques are commonly employed to derive models directly from data, enabling predictions and insights into sample behavior. Machine Learning (ML) has emerged as a powerful tool in this domain, as it can automatically identify patterns and perform inference with minimal human intervention. Applications of ML in cosmology include, for example, \citep{chacon2021classification,arjona2021testing,von2021inferring,carlos2023,2024PhyS...99k5007P,mukherjee2025newsim5sigmatension,borghetto2025boundeddarkenergy}. A natural question that arises is whether the age of the Universe can be reconstructed using a machine learning framework.

This work aims to reconstruct galaxy ages using ML, employing synthetic data generated through the Monte Carlo Method (MCM) \citep{metropolis1987beginning} as a basis for training and analysis. The structure of this paper is as follows: Section~\ref{Theoretical} introduces the cosmological background, while Section~\ref{SL} presents the machine learning techniques used. Section~\ref{Data} discusses the observational and simulated data, along with the statistical methods employed. In Section~\ref{results}, we analyze the performance of the ML models in constraining cosmological parameters such as $\omega$ and $\Omega_m$, assessing both their predictive power and consistency with the fiducial cosmology. Finally, Section~\ref{conclusion} summarizes our findings and outlines future directions.

\section{Theoretical framework}\label{Theoretical}
\subsection{Age of the galaxies}\label{AoO}

The theoretical age-$z$ relation $t(z_i)$ of an object at redshift $z_i$ can be written as  \cite{1988Sandage,1993peebles} 
\begin{equation}\label{eq:t_0} 
t(z_i,\textbf{p}) = \int_{0}^{\infty} \frac{dz'}{(1+z') \mathcal{H}(z',\textbf{p})}\;, 
\end{equation}
in which $\textbf{p}$ stands for the parameters of the cosmological model under consideration and $\mathcal{H}(z', \textbf{p})$, is the normalized
 Hubble parameter, given by
\begin{equation}\label{Hz}
    \mathcal{H}(z',\textbf{p})=\left[\sum_i\Omega_i(1+z)^{3(1+\omega_i)} + \Omega_k(1+z)^2 \right]^{\frac{1}{2}}\;,
\end{equation}
with $\Omega_i$ $(i=m,r,x)$ describing the density parameter of the $i$-th component, and $\Omega_{k}$ is the curvature parameter. From the observational viewpoint, the total age of a given  galaxy at redshift $z$ is
\begin{equation}
    t(z)^\text{obs}=t_G(z)+ \tau(z)\;.
\end{equation}
The quantity $t_G(z)$ represents the estimated age of the oldest stellar population, while $\tau(z)$ denotes the incubation time, or delay factor. This parameter accounts for our lack of knowledge regarding the time elapsed from the beginning of structure formation in the Universe to the formation time, $t_f$, of the object of interest (for a more detailed discussion, see, e.g., \cite{capozziello2004constraining,pires2006lookback}).

\subsection{Deceleration parameter}

We need a quantity to quantify the Universe's expansion rate, and the deceleration parameter gives this. It can be written as a function of $z$ and the density parameters. Differentiating equation~\eqref{Hz} and  using the definition  $q_0 \equiv -\ddot{a}a/\dot{a}^2$ \cite{1993peebles}, we arrive at 

\begin{equation}
   q(z)= \frac{3}{2}\frac{\sum_i\Omega_i(1+\omega_i)(1+z)^{3(1+\omega_i)}}{\sum_i \Omega_i(1+z)^{3(1+\omega_i)} + \Omega_k(1+z)^2 } -1 \;. 
\end{equation}
Note that for $z=0$, we obtain the actual deceleration parameter, a positive value of $q_0$ meaning that the universe's expansion is decelerating, while with a negative value, we have an accelerating universe.

\section{Supervised Learning}\label{SL}

Supervised learning is the most widely used form of machine learning, aimed at learning a mapping between inputs, $x_i$, and outputs, $y_i$, given a dataset $ \mathcal{D} = \{(x_i, y_i)\}_{i=1}^N $, where $\mathcal{D}$ is the training set, and $N$ is the number of training examples. The input $x_i$ is typically a D-dimensional vector representing attributes like the color or size of a shirt, but can also represent complex structures such as images or time series.

The outputs $ y_i $ can be categorical (e.g., the spectral type of a star) or continuous (e.g., galaxy age). When $y_i$ is categorical, the problem is classification, and when $y_i\in\mathbb{R}$, it is regression.

In both cases, we assume $y=f(x)$ for an unknown function $f$. Given a training set, we estimate $f$ and predict $ \hat{y} = \hat{f}(x) $ for new inputs (test set). The choice of training and test set sizes is crucial, typically using 70\% for training and 30\% for testing.

In more complex scenarios, it’s useful to return a probability distribution, $ p(y\mid x,\mathcal{D})$, or more specifically, $ p(y\mid x,\mathcal{D},M) $, where $ M $ is the model. The best estimate for the true labels can be obtained using the Maximum A Posteriori (MAP) estimate
\begin{equation}
    \hat{y} = \hat{f}(\mathbf{x}) = \text{argmax} \, p(y \mid x, \mathcal{D})\;.
\end{equation}

Inputs can also vary in measurement types, leading to distinctions in prediction methods. Some methods are more suited for quantitative inputs, others for qualitative, and some for both. The following sections will discuss various supervised learning regression methods, with terms as they appear in the \texttt{Scikit-learn}\footnote{\url{https://scikit-learn.org/stable/}}
library highlighted.

 Evaluating the quality of ML methods and optimizing them requires error analysis and hyperparameter tuning. Accuracy, measured as the fraction of correct predictions, is a primary metric, while the coefficient of determination\footnote{\url{https://scikit-learn.org/stable/modules/generated/sklearn.metrics.r2_score.html}} ($R^2$) provides a measure of fit quality. Cross-validation techniques, such as K-Fold\footnote{\url{https://scikit-learn.org/stable/modules/generated/sklearn.model_selection.KFold.html}} and ShuffleSplit\footnote{\url{https://scikit-learn.org/stable/modules/generated/sklearn.model_selection.ShuffleSplit.html}}, are employed to prevent overfitting and assess the model's generalization ability. Hyperparameter optimization, essential for enhancing algorithm performance, is conducted using tools like GridSearchCV\footnote{\url{https://scikit-learn.org/stable/modules/generated/sklearn.model_selection.GridSearchCV.html}}, which systematically explores parameter combinations to identify the best model configuration.

\subsection{Classification and Regression Trees (CART)}
Classification and Regression Trees (CART) are used for both classification and regression tasks. The CART algorithm builds a tree structure to make decisions based on input features. We
evaluate the algorithm hyperparameter values that
best fit the input simulations through a grid search.
Hence, our grid search over the CART hyperparameters are given by the following inputs:
\small{
\begin{lstlisting}
GridSearchCV(estimator=
            DecisionTreeRegressor(),
            param_grid={
            `ccp_alpha': [0.01, 
            0.009, 0.008, 0.001],
            `criterion': (`squared_error',
            `friedman_mse',`absolute_error'),
            `max_depth':range(1,31),
            `min_samples_leaf':range(1,6)})
\end{lstlisting}}
\subsection{Support Vector Regression (SVR)}
Support Vector Regression (SVR) is a type of regression that uses support vector machines. It aims to find a function that deviates from the actual observed values by a margin no greater than a specified threshold. The components of the SVR hyperparameter grid search are:
\small{
\begin{lstlisting}
GridSearchCV(estimator=SVR(),
            param_grid={
            `C': [0.01, 0.05, 
            0.08, 0.02, 0.1, 0.2,
            0.3, 0.04],
            `coef0': [0.1, 0.2, 0.6],
            `epsilon': [0.01, 0.05, 
            0.08, 0.02, 0.1, 0.2, 
            0.3,0.4], 
            `gamma': (`auto', `scale'),
            `kernel': (`linear', `poly')})   
\end{lstlisting}}

\subsection{Multilayer Perceptron Regression (MLPR)}

Multilayer Perceptron Regression (MLPR) is a type of artificial neural network used for regression tasks. It consists of an input layer, one or more hidden layers, and an output layer. Using a grid search, we determine which algorithm hyperparameter values best fit the input simulations. Thus, the following inputs provide our grid search across the MLPR hyperparameters:
\small{
\begin{lstlisting}
GridSearchCV(cv=ShuffleSplit(n_splits=5, 
            random_state=None,
            test_size=None,
            train_size=None),
            estimator=MLPRegressor(),
            param_grid={`activation':
            (`identity',`relu'),
            `alpha': [0.01, 0.02,
            0.04, 0.05],
            `max_iter': [10000, 20000,
            30000, 40000],
            `solver': (`lbfgs', `sgd',
            `adam')},
            return_train_score=True,
            scoring=`r2')
\end{lstlisting}}


\section{Overview of Data and Statistical Evaluation}\label{Data}
\begin{figure}
    \centering
    \includegraphics[width=0.5\textwidth]{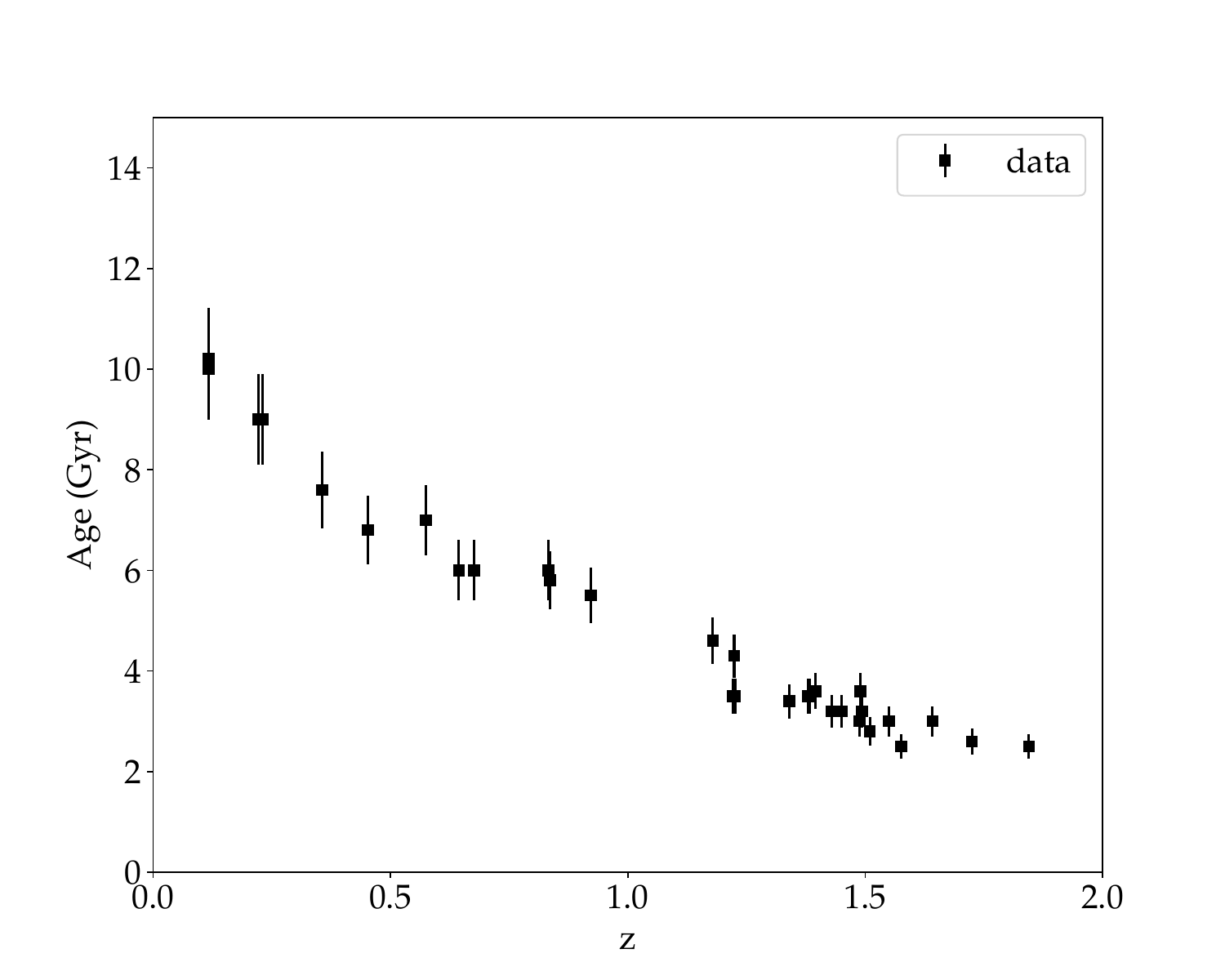}
    \caption{Original sample of 32 galaxies, as described in the dataset, covering the redshift range $0.11 \leq {z} \leq 1.84$ from \cite{simon2005constraints}, with a $10\%$ uncertainty in the measurements.}
    \label{fig:age-data}
\end{figure}

For the ages $t(z_i)$, we will utilize a dataset consisting of ages from 32 galaxies distributed across the redshift range $0.11 \leq {z} \leq 1.84$ \citep{simon2005constraints}. This dataset is divided into three sub-classes: 10 early-type field galaxies \citep{savage2005age,doran2007baryon}, whose ages were derived using SPEED models \citep{komatsu2009five}; 20 red galaxies from the Gemini Deep Deep Survey (GDDS) \citep{abraham2004gemini}; and two radio galaxies, LBDS 53W091 and LBDS 53W069 \citep{percival2010baryon,nolan2001sun}. These observations carry an uncertainty of $10\%$, see the Figure~\ref{fig:age-data}.

\subsection{Simulated Data\label{sec
}}

To obtain our simulated samples of $t(z)$, we used the Monte Carlo Method (MCM) \cite{metropolis1987beginning} and adopted a Gaussian distribution centered on a flat fiducial $\Lambda$CDM model with $\Omega_m = 0.315 \pm 0.007$ and $H_0 = 67.4\pm 0.5$ km s$^{-1}$ Mpc$^{-1}$, values consistent with the best fits obtained by \cite{aghanim2020planck}, such that the standard deviation of this probability distribution corresponds to the percentage error of the $t(z)$ determinations.

According to information provided by observational missions, such as the Atacama Cosmology Telescope (ACT) in Chile \cite{kosowsky2003atacama} and the Southern African Large Telescope (SALT) in South Africa \cite{stobie2000design}, a sample of up to 2000 old galaxies observed within a range of $0 < z < 1.5$ is expected, with $10\%$ uncertainty in their age determinations. Although such a large sample of galaxy ages is not yet available, we consider this a hypothetical scenario to evaluate the robustness of the supervised ML techniques used in this work—namely, CART, MLPR, and SVR. Therefore, to simulate this idealized case, we generate samples of 100, 1000, 2000, 3334, and 6680 galaxies, assuming a $10\%$ Gaussian error in age estimates. The sizes 3334 and 6680 correspond to test cases where the predicted sample from ML models represents $30\%$ of the total, thus mimicking the full data scenario.

\begin{figure}[!ht] 
\centering 
\includegraphics[width=0.5\textwidth]{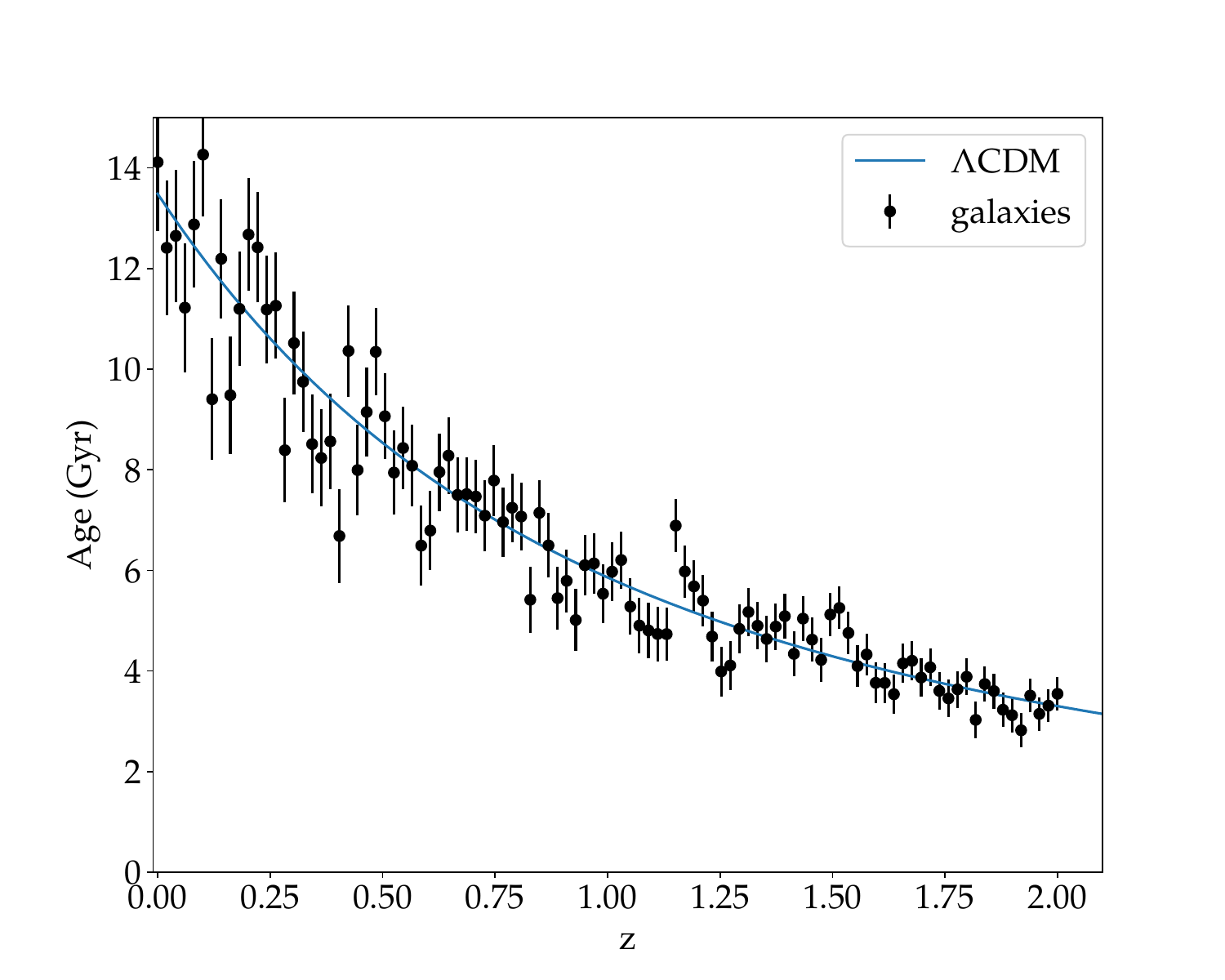} 
\caption{Simulated age sample with 100 points.} 
\label{fig:age-simu} 
\end{figure}

The Figure~\ref{fig:age-simu} shows an example of simulated ages for a sample with 100 points, where the age is measured in billions of years (Gyrs). An error of $\sigma_t = 10\%$ is adopted, and the solid line corresponds to the fiducial model on which the simulations are based.

\subsection{Statistical Data Analysis}

To estimate the best fit for our regression, we will use the minimum chi-squared test, defined as

\begin{equation}\label{chi} 
\chi^2\equiv\sum_i\frac{[t_{G}- t_{predicted}]^2}{\sigma_T^2}\;, 
\end{equation} 
where $\sigma_T^2 = \sigma_o^2 + \sigma_s^2$ corresponds to the propagated error given by the quadratic sum of the observational and simulation errors. It is important to note that in equation~\eqref{chi}, the sum will always yield a non-negative result, as the sum is quadratic. This implies the following consequence: if the sum is zero, it indicates high precision, as the expected value matches the observed value. Conversely, if the sum is large, it suggests that the result is far from expected, indicating that the model does not describe the data satisfactorily.

We can use the minimum chi-squared solution to write the maximum likelihood function, given by 
\begin{equation} \ln \mathcal{L}= -\frac{\chi^2}{2}\;. 
\end{equation} 

Next, we need to describe the prior function, $p(\theta)$, which encodes any prior knowledge we have about the parameters. After completing this procedure, we can obtain our sample distribution using \texttt{emcee} \citep{emcee2013}, and then obtain our contour regions using \texttt{ChainConsumer} \citep{Hinton2016}.

In order to quantify the variance of the reconstructed universe age as a function of redshift, we adopt the standard decomposition of the mean squared error

\begin{equation}\label{eq:BVT}
    \text{BVT}(z) =  \left\langle\Delta t(z)^2 \right\rangle - \left\langle\Delta t(z)\right\rangle^2,
\end{equation}
where $\Delta t(z) = t^{\text{pred}}(z) - t^{\text{fid}}(z)$ represents the difference between the predicted and fiducial values of the universe age at a given redshift $z$. The first term corresponds to the mean squared error (MSE), while the second term is the squared bias. This decomposition allows us to identify whether the observed reconstruction errors are dominated by statistical fluctuations (variance) or by systematic deviations (bias).

\section{Results}\label{results}
In this section, we present our main results derived from the methodology used in this study. As previously described in Section 3, we split 70\% of the simulated sample for the training set and 30\% for the test set, resulting in test samples with 30, 300, 600, 1001, and 2004 data points, corresponding to the simulated galaxy age samples of 100, 1000, 2000, 3334, and 6680 data points (Section 4.2). In our analyses, the number of points in the prediction (reconstruction) will match the number of points in the test set.

\begin{figure*}[!ht]
    \centering
    \includegraphics[width=
    \linewidth]{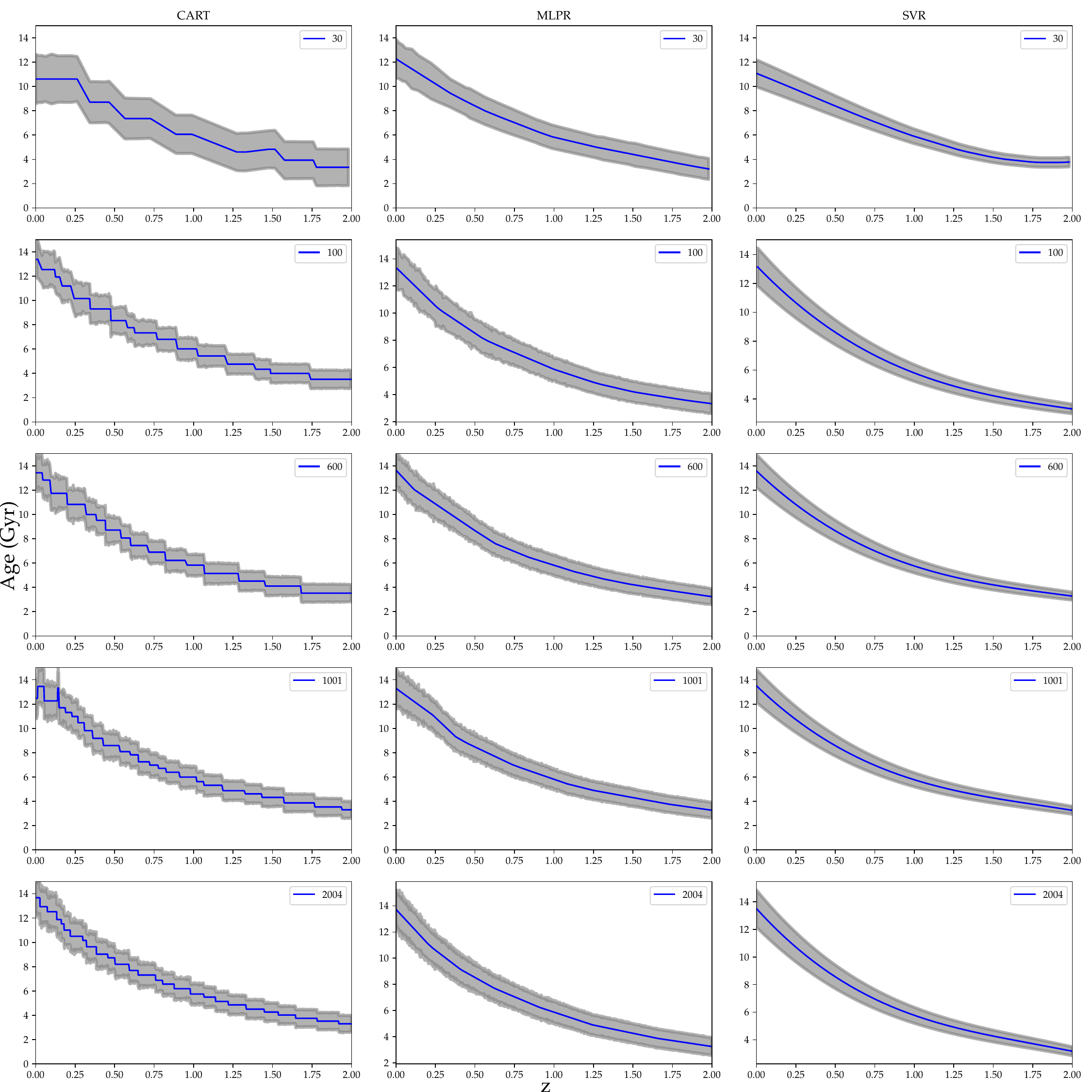}
    \caption{Reconstructions via ML for five samples using three techniques: CART (first column), MLPR (second column), and SVR (third column). The solid blue line and the gray region represent the age reconstruction and the $1\sigma$ error for each sample, respectively.}
    \label{fig:allresuts}
\end{figure*}

In Figure \ref{fig:allresuts}, we illustrate the ML-based reconstructions for the five samples using the three different techniques adopted in this work: CART (1st column), MLPR (2nd column), and SVR (3rd column). In all graphs, the x-axis corresponds to the predicted age, and the y-axis corresponds to the redshift. The solid blue line and the gray region represent the age reconstruction and the 1$\sigma$ error for each sample, respectively. Note that the overall behavior is similar across all techniques. However, we can observe some nuances specific to each analysis
\begin{itemize}
    \item For all techniques, as we increase the number of points, the age of galaxies decreases at high redshifts, and the dip becomes more pronounced. We cannot fully explain why this latter behavior occurs, as each technique has its own peculiarities;
    \item In CART, the step-like behavior reflects the characteristic of decision tree techniques rather than the sample itself, and it also shows the highest reconstruction error;
    \item Note also that for all techniques, the propagated error decreases as we increase the number of sample points. For SVR, this reduction is even more pronounced compared to the other two techniques, because the MSE for SVR is approximately 10 times smaller than the MSE for the other two techniques.
\end{itemize}

\begin{figure}[!ht]
    \centering
    \includegraphics[width=0.5\textwidth]{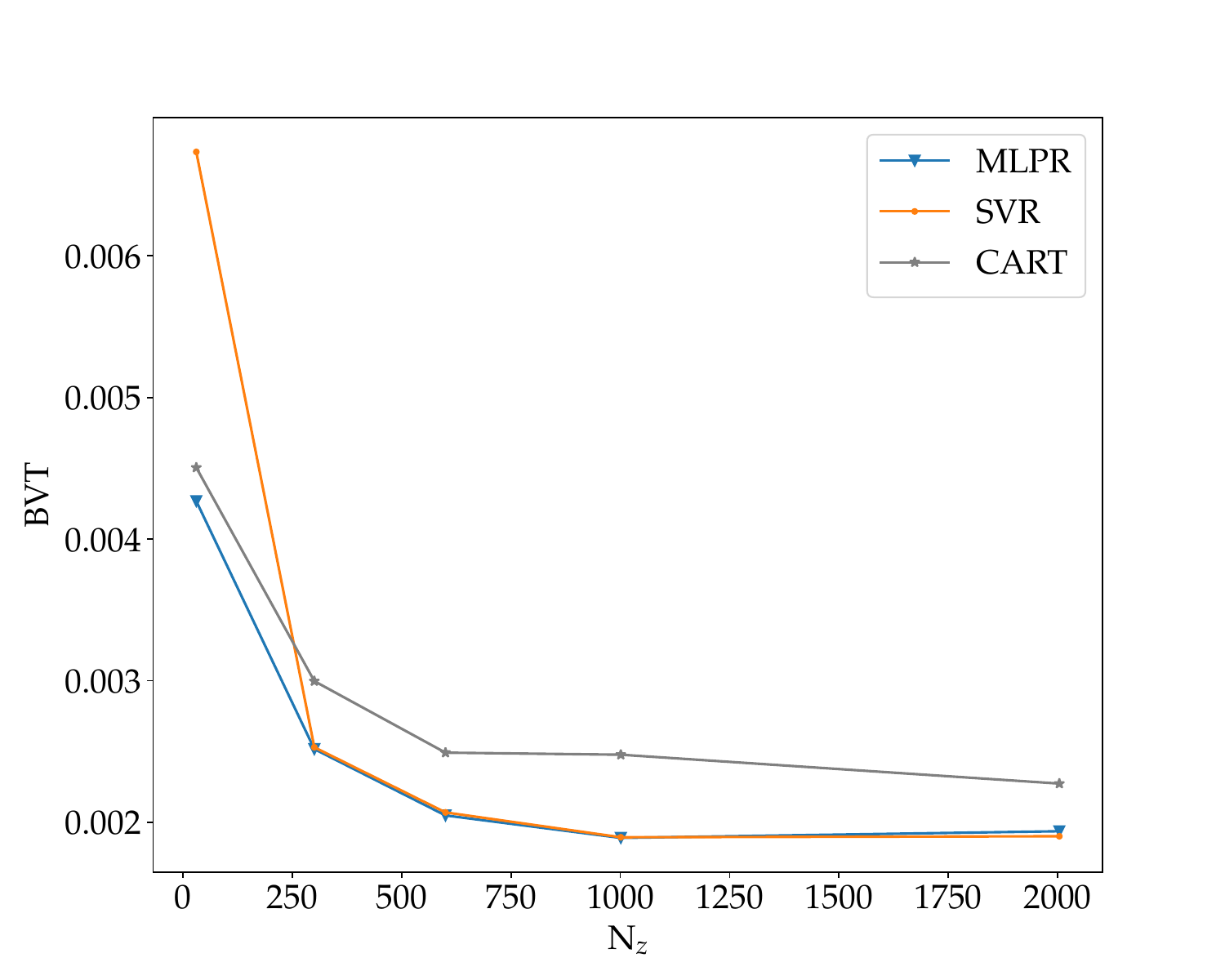}
    \caption{Bias-variance tradeoff (BVT) as a function of the number of redshift points $N_z$ for different machine learning regressors. All models exhibit decreasing variance with increasing data size, with SVR showing the lowest BVT overall.}
    \label{fig:BVT}
\end{figure}

Using the bias-variance decomposition defined in equation~\eqref{eq:BVT}, we compute the variance component of the reconstruction error for each algorithm as a function of the number of redshift sampling points $N_z$. The results are shown in Figure~\ref{fig:BVT}. As expected, all methods display a decreasing trend in BVT as $N_z$ increases, reflecting the reduction of stochastic fluctuations in the predictions due to larger training samples. Among the models analyzed, SVR exhibits the lowest BVT values across all data regimes, indicating a better balance between bias and variance. In contrast, CART presents a higher residual variance, especially for larger $N_z$, which may be attributed to the piecewise nature of its decision rules. These results reinforce the robustness of SVR in reconstructing the age of the universe from simulated cosmological data.

Based on the reconstructions for each technique illustrated above, we were able to find the best fits for the free parameters of our adopted cosmological model, the flat geometry $\omega$CDM dark energy model.

\begin{figure*}[!ht]

{\includegraphics[width=0.3\textwidth]{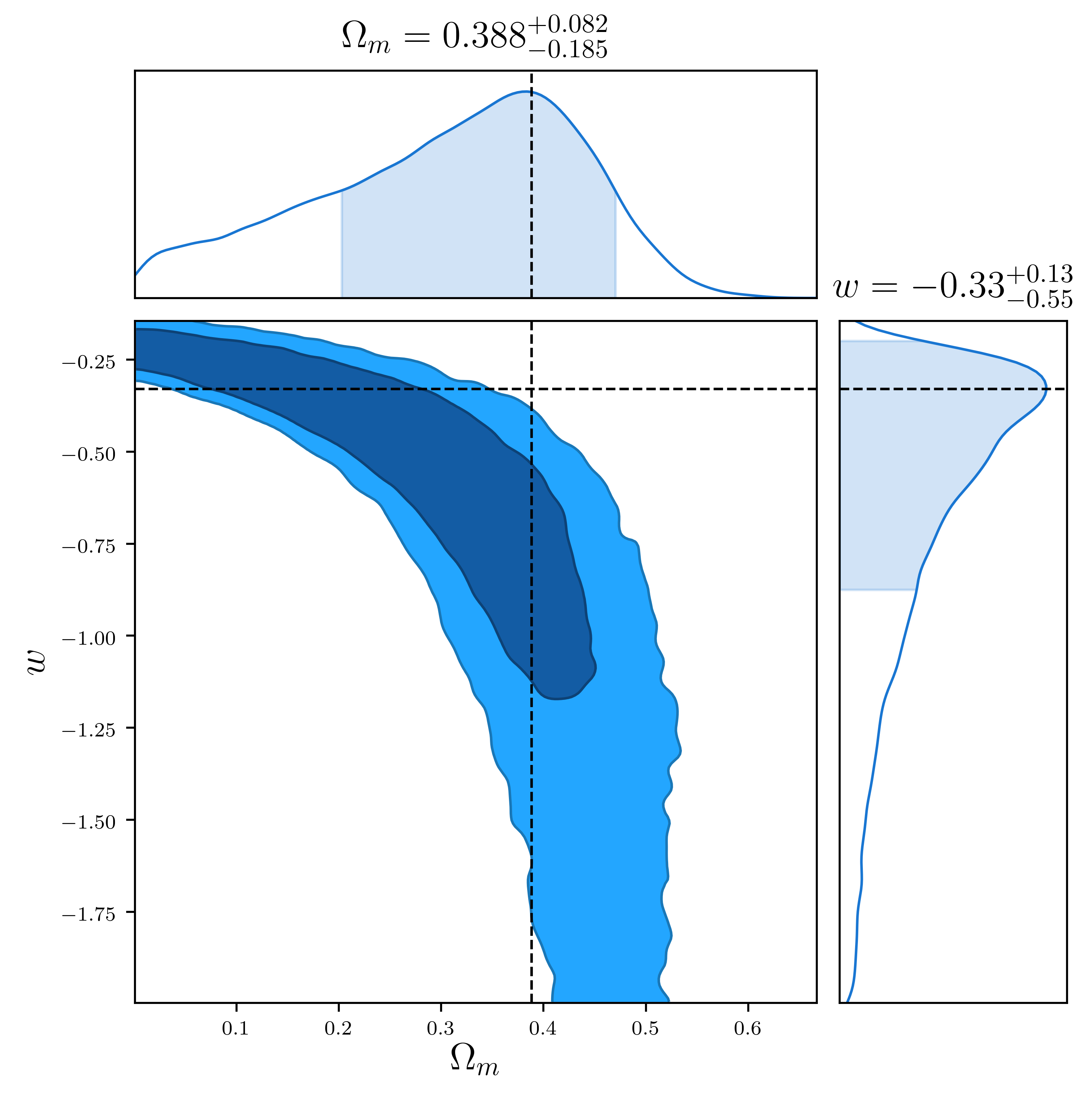}} 
{\includegraphics[width=0.3\textwidth]{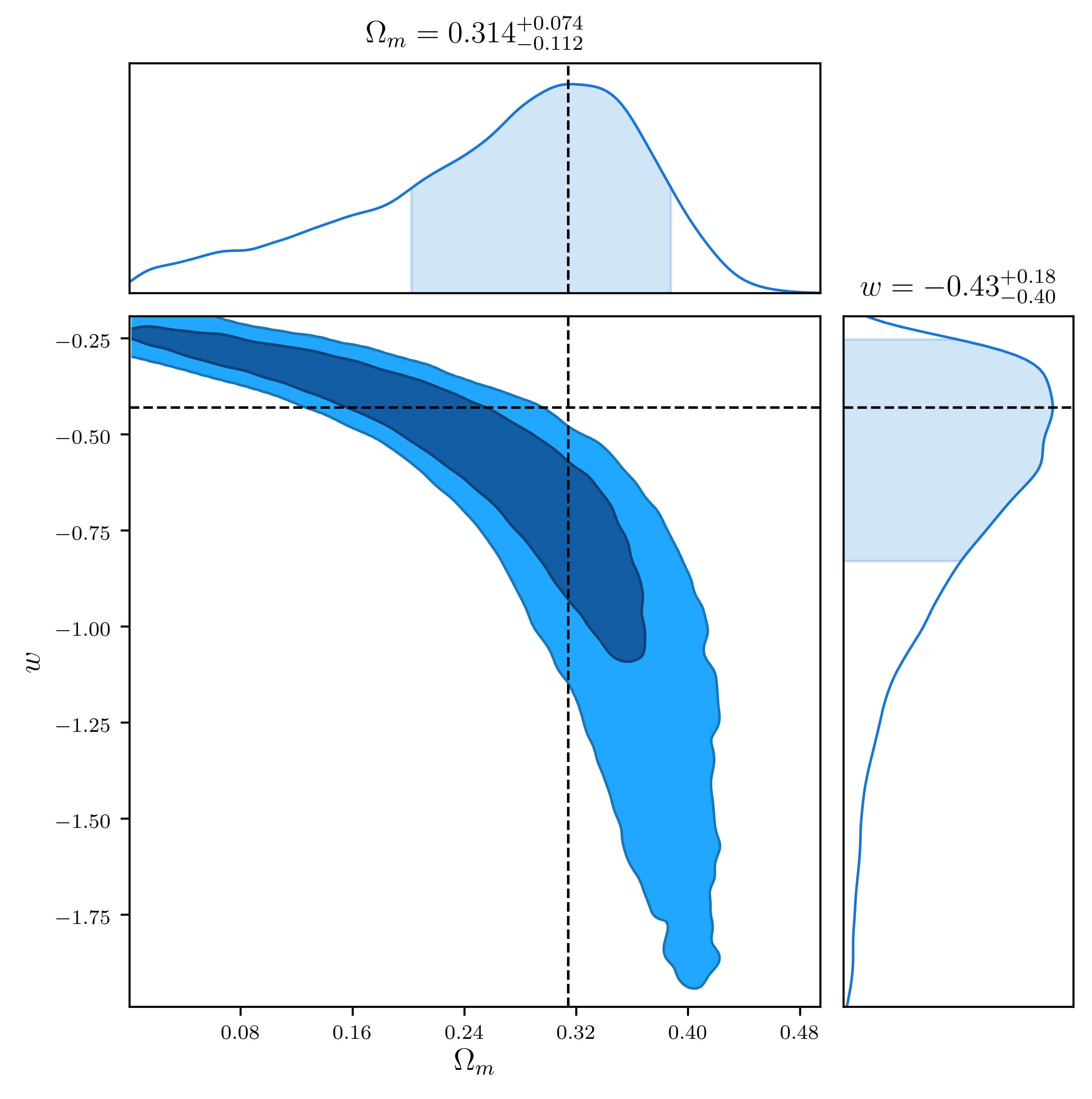}} 
{\includegraphics[width=0.3\textwidth]{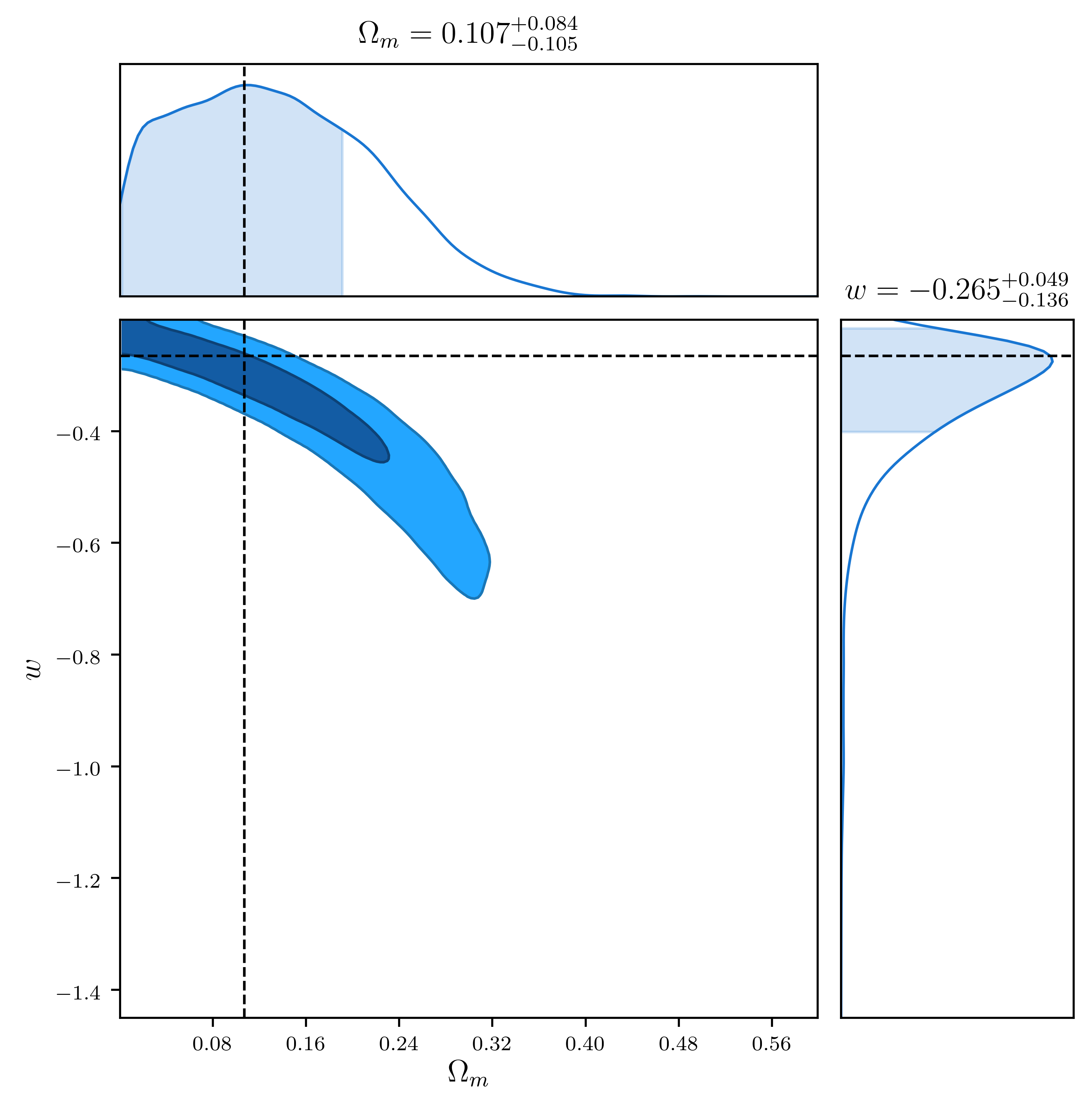}}
\caption{Results for the $\omega$CDM${k=0}$ model with 30 predicted age points using the CART, MLPR, and SVR techniques, respectively. The $68.3\%$ and $95.4\%$ confidence regions are shown in the $\Omega_m - \omega$ plane. The dashed lines represent the best fits illustrated in the figures with their respective probability distributions.}
\label{fig:30}
\end{figure*}

\begin{figure*}[!ht]
{\includegraphics[width=0.3\textwidth]{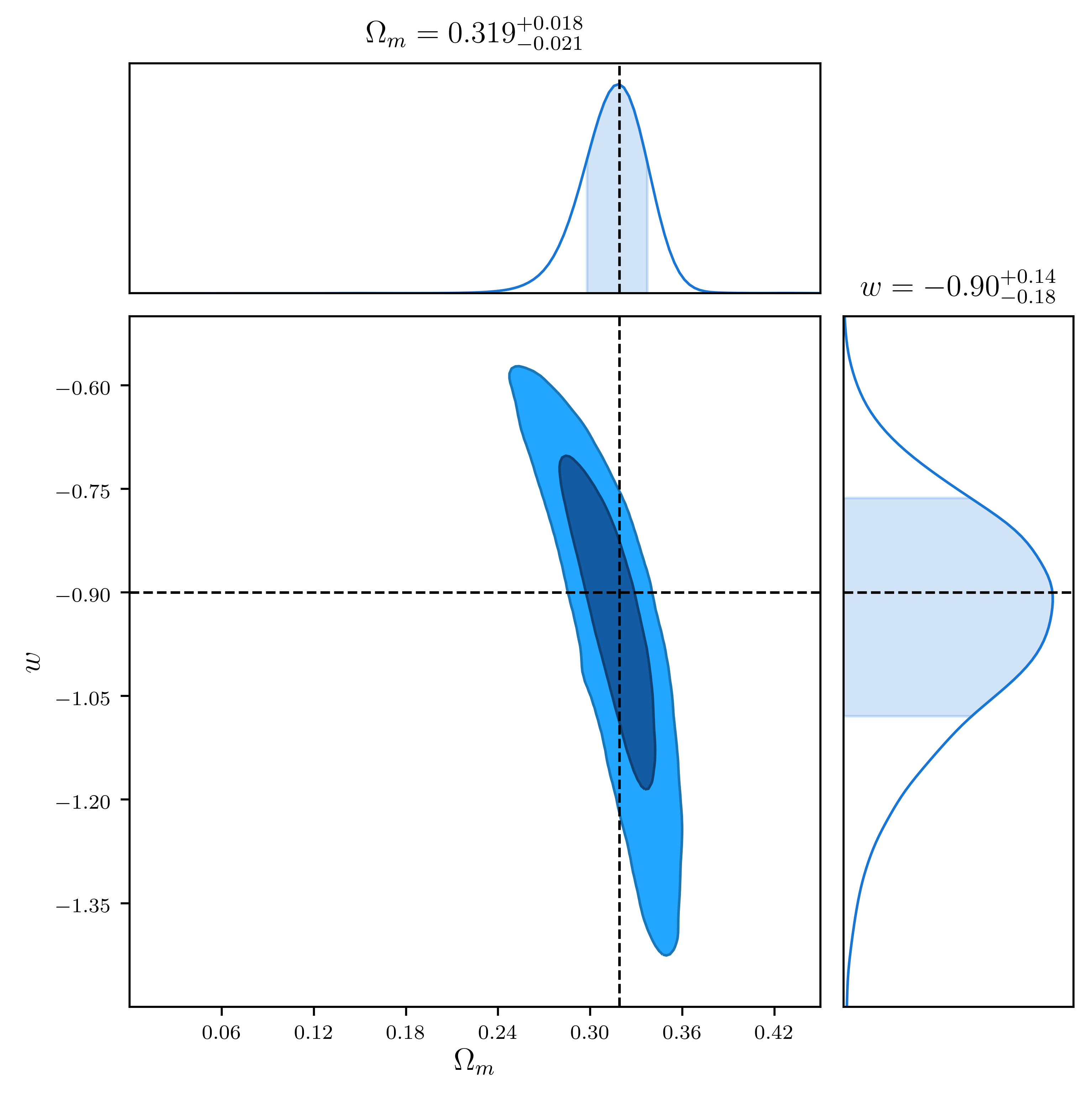}}
{\includegraphics[width=0.3\textwidth]{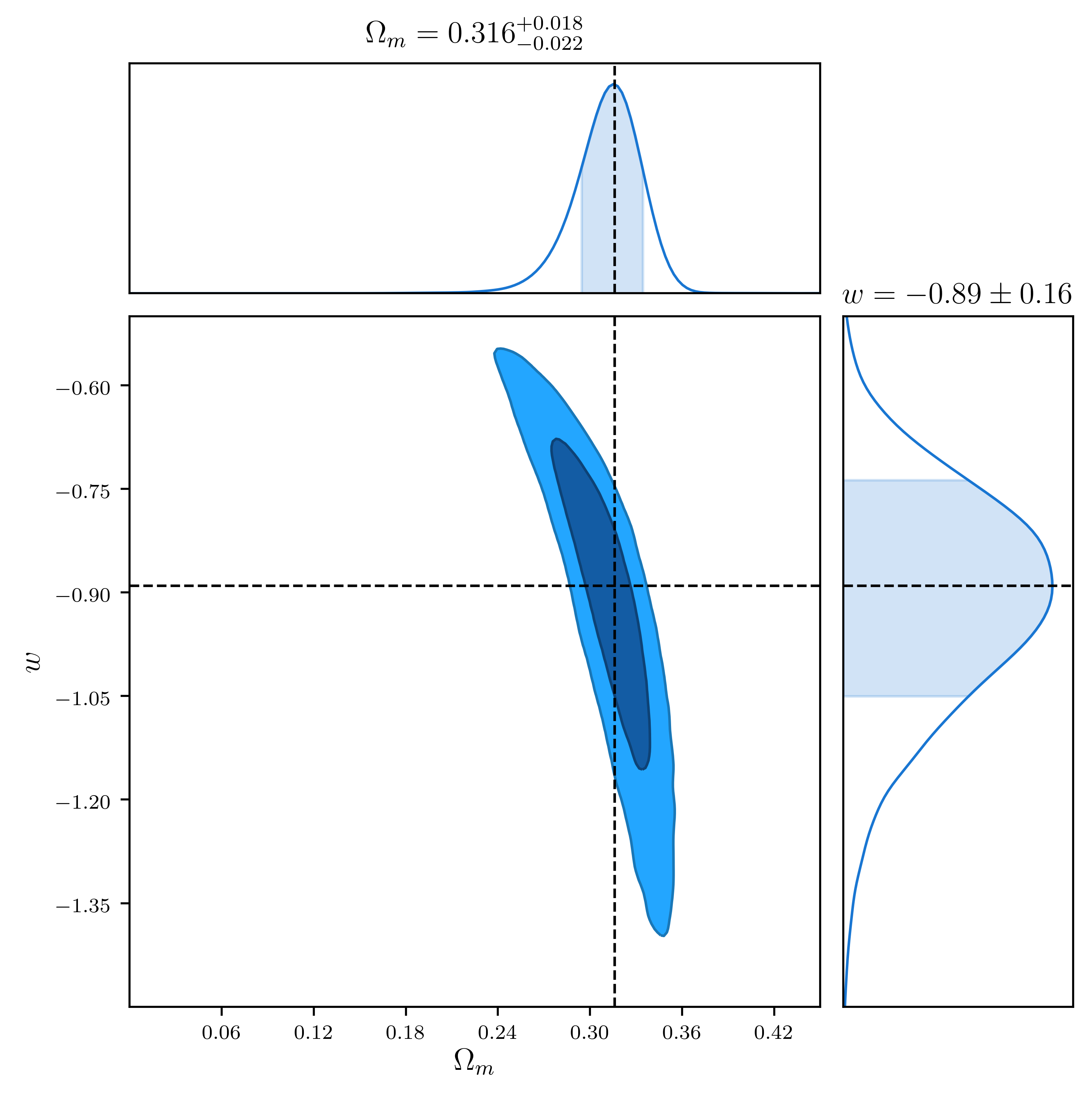}}
{\includegraphics[width=0.3\textwidth]{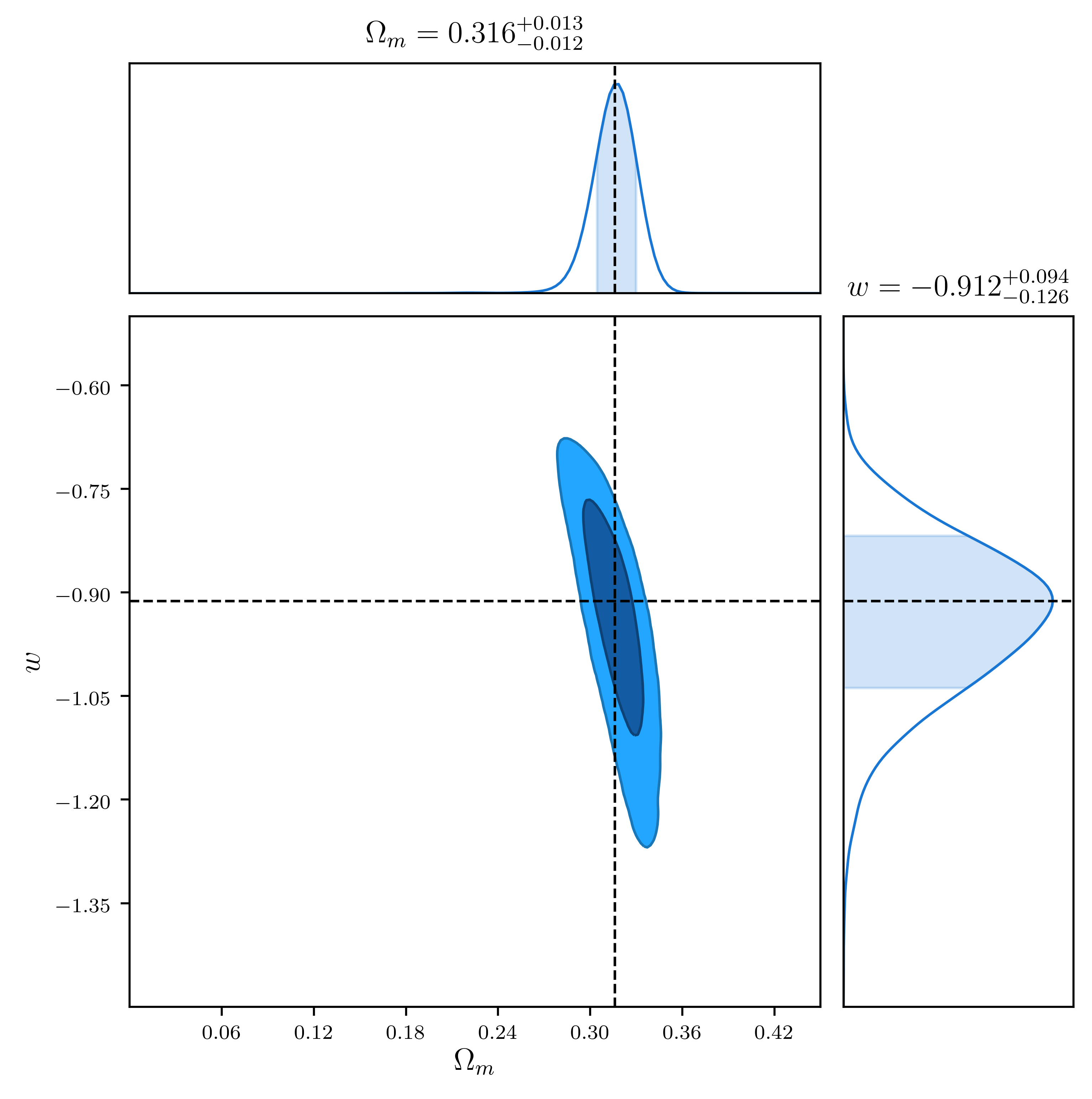}}
\caption{Same as Figure \ref{fig:30}, but with 300 predicted ages.}

\label{fig:300}
\end{figure*}

\begin{figure*}[!ht]
{\includegraphics[width=0.3\textwidth]{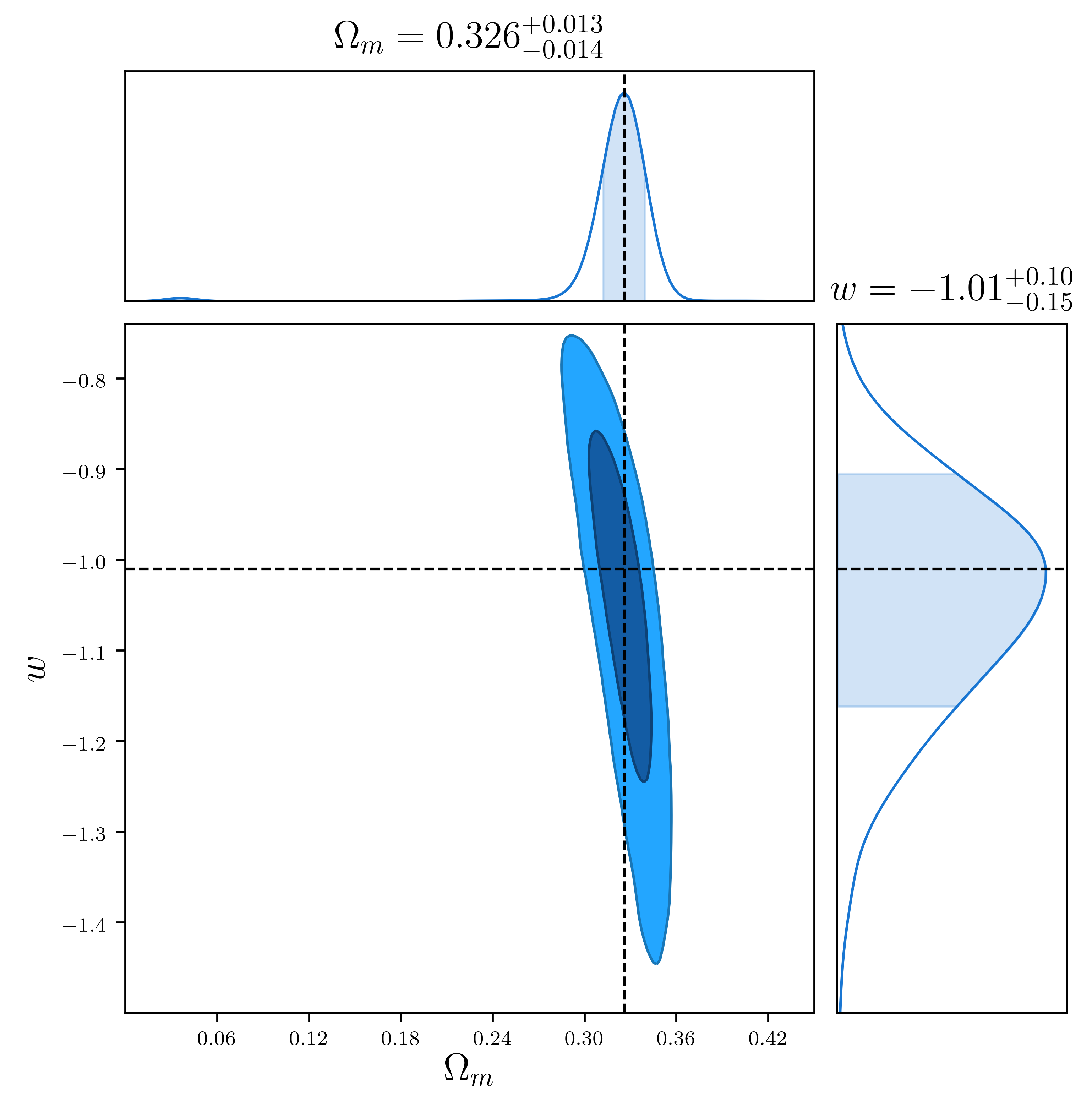}}
{\includegraphics[width=0.3\textwidth]{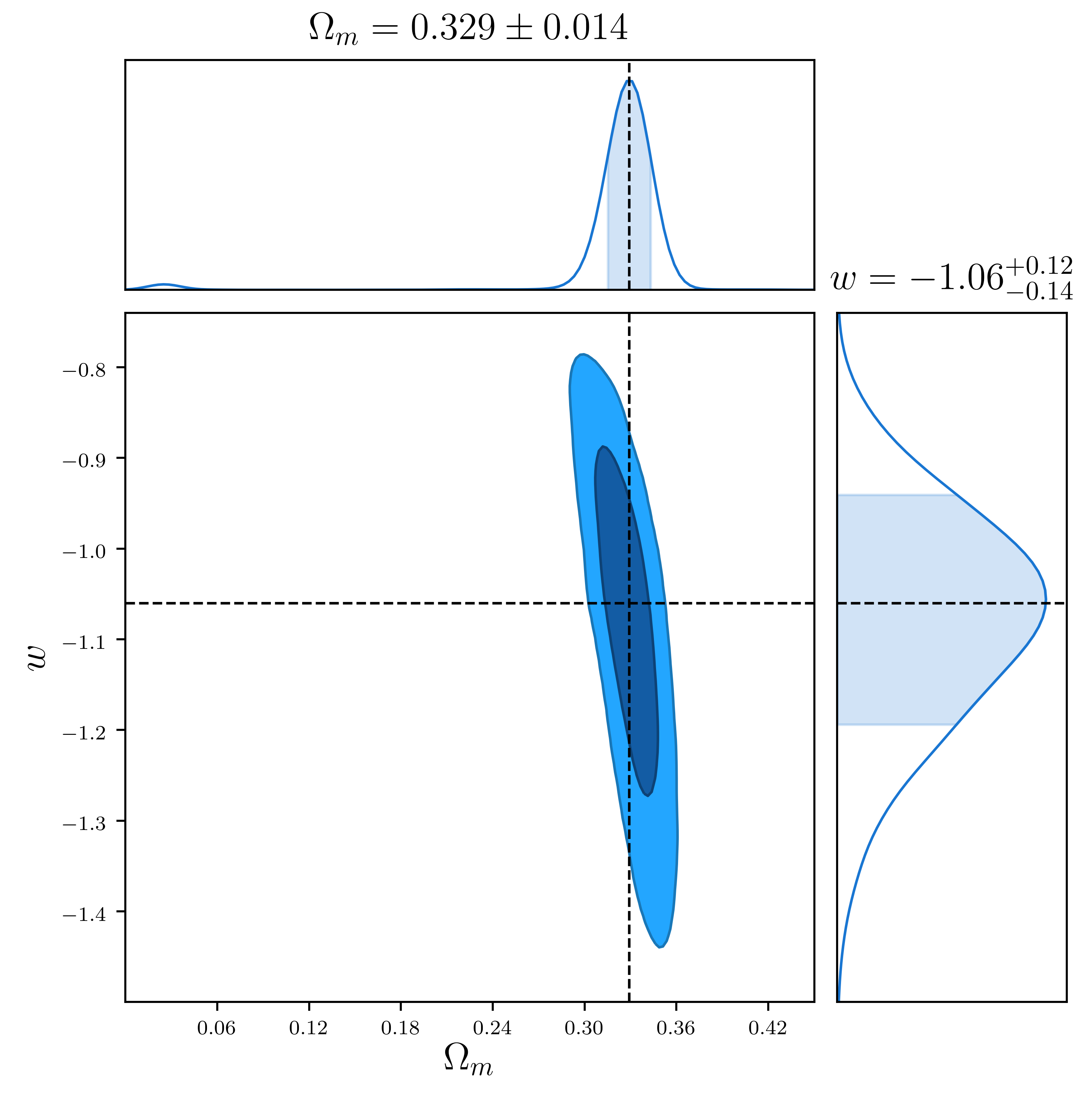}}
{\includegraphics[width=0.3\textwidth]{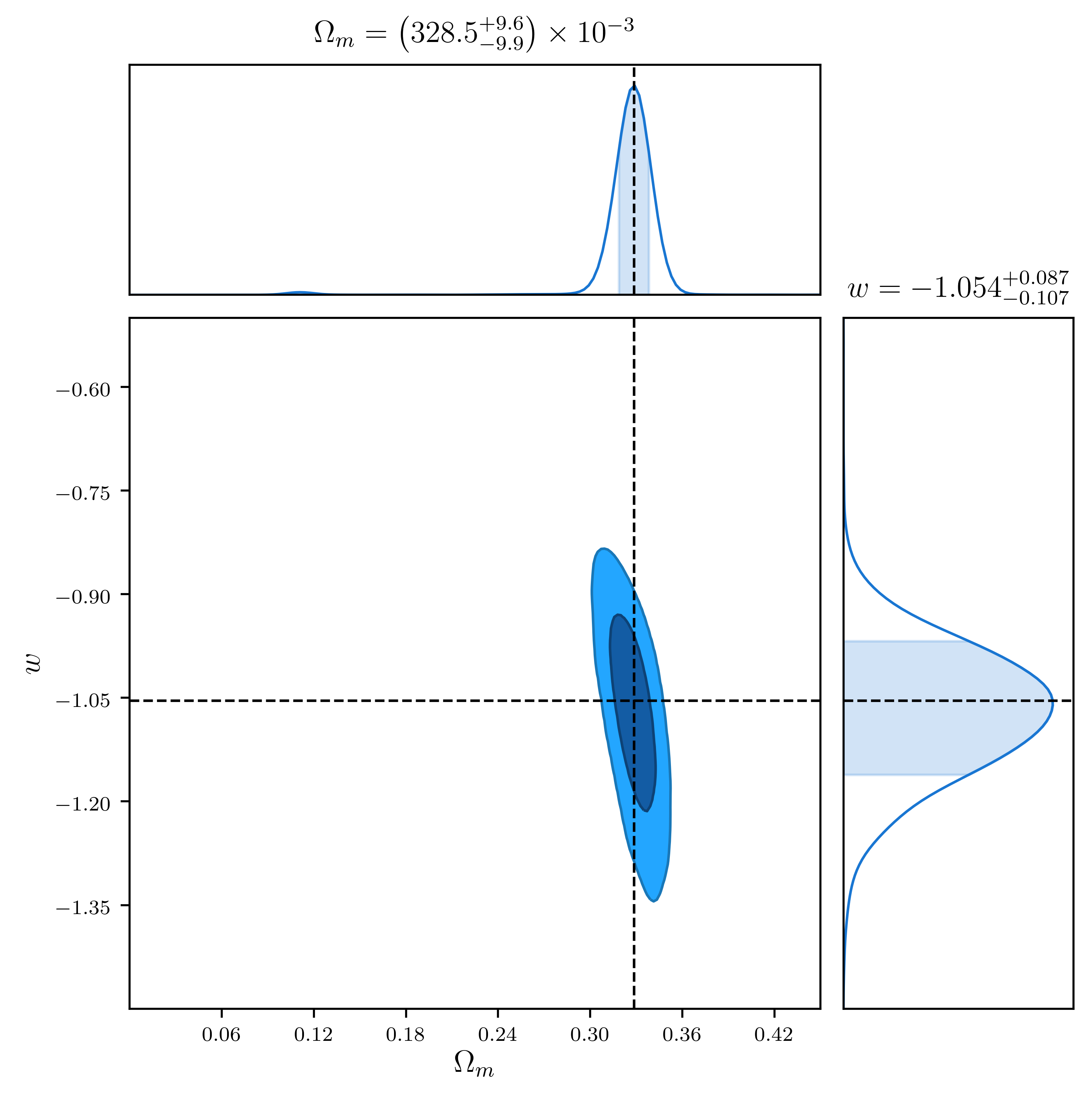}}
\caption{Same as Figure \ref{fig:30}, but with 600 predicted ages.}
\label{fig:600}
\end{figure*}

\begin{figure*}[!ht]
{\includegraphics[width=0.3\textwidth]{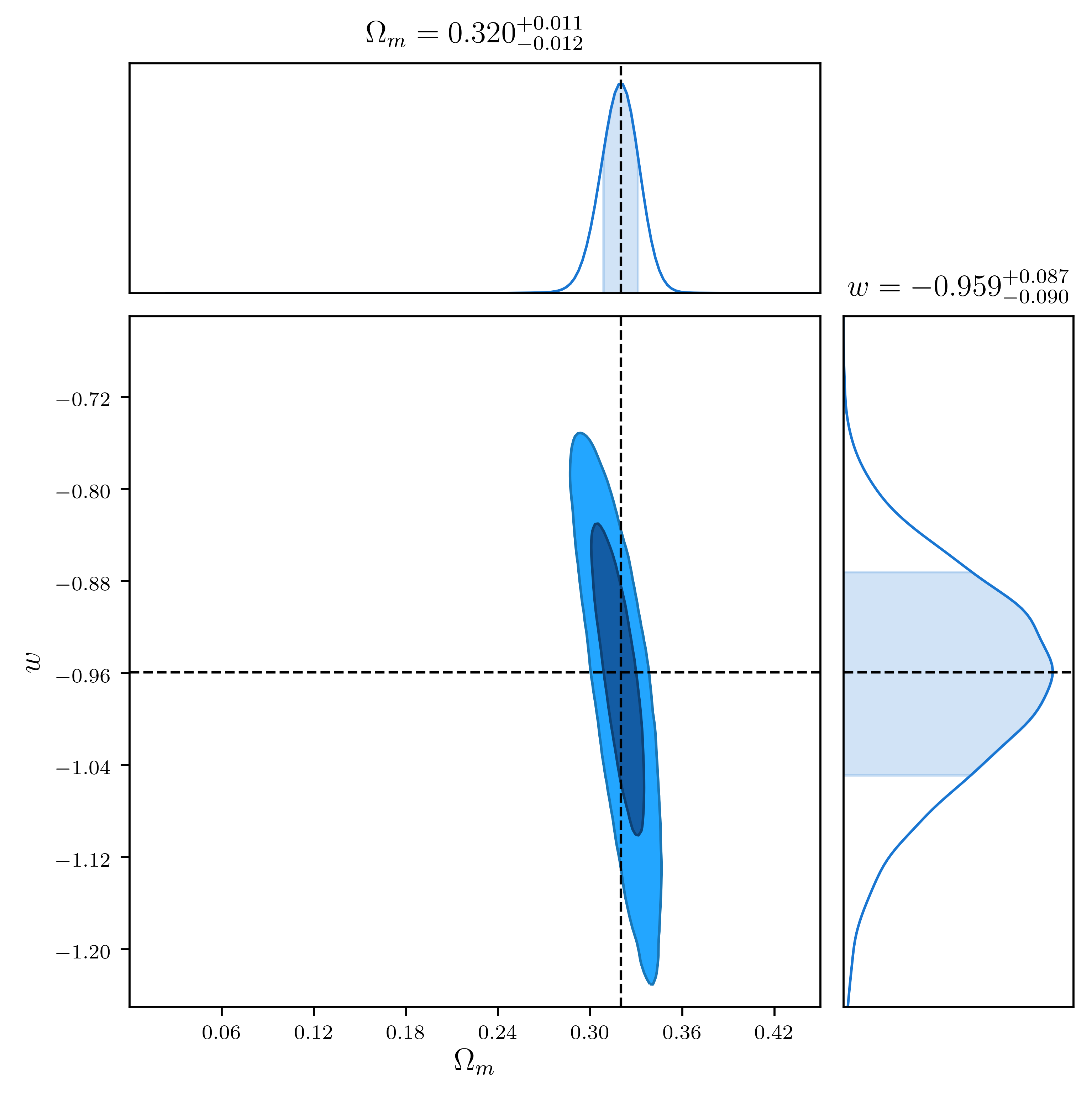}}
{\includegraphics[width=0.3\textwidth]{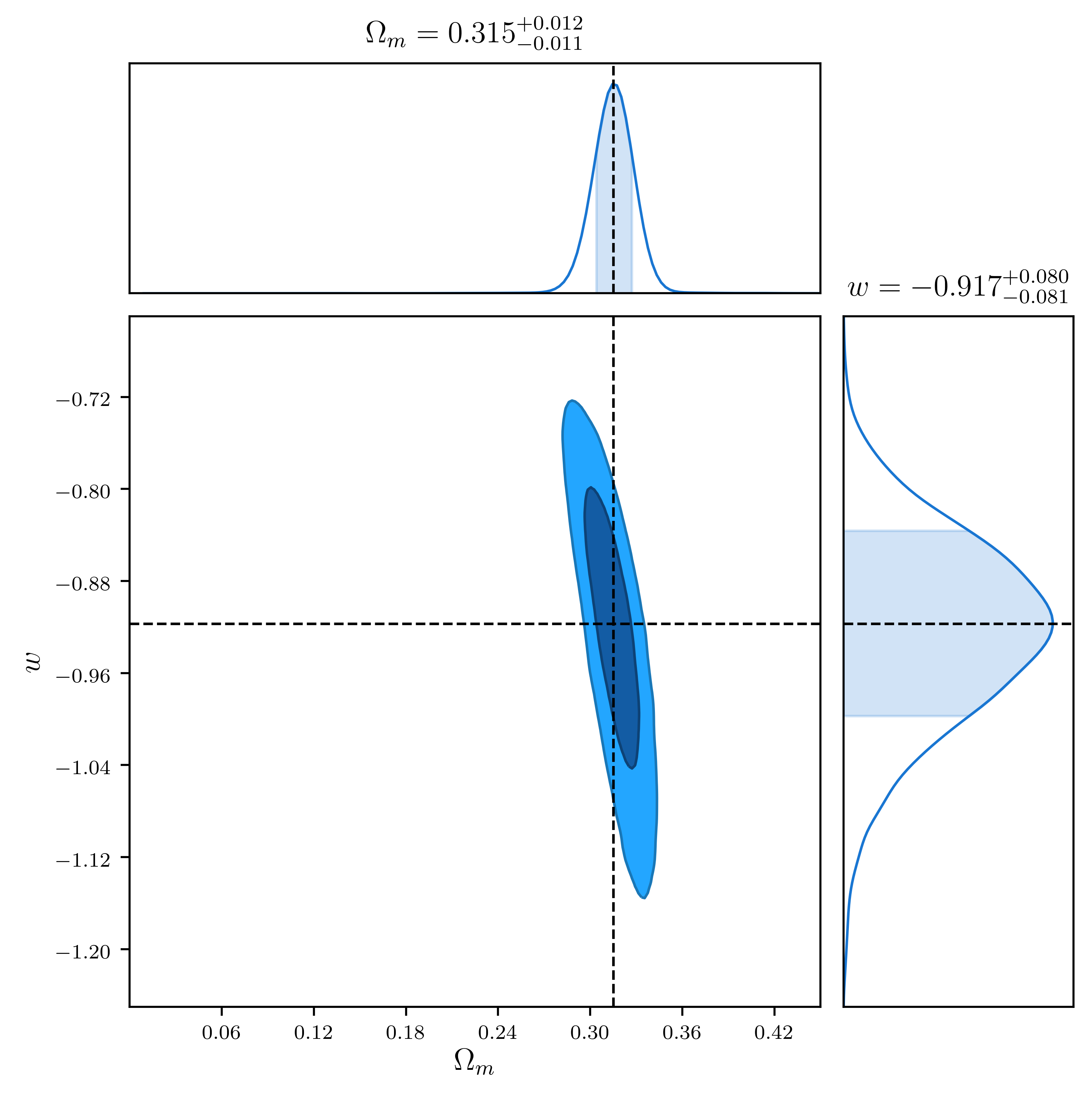}}
{\includegraphics[width=0.3\textwidth]{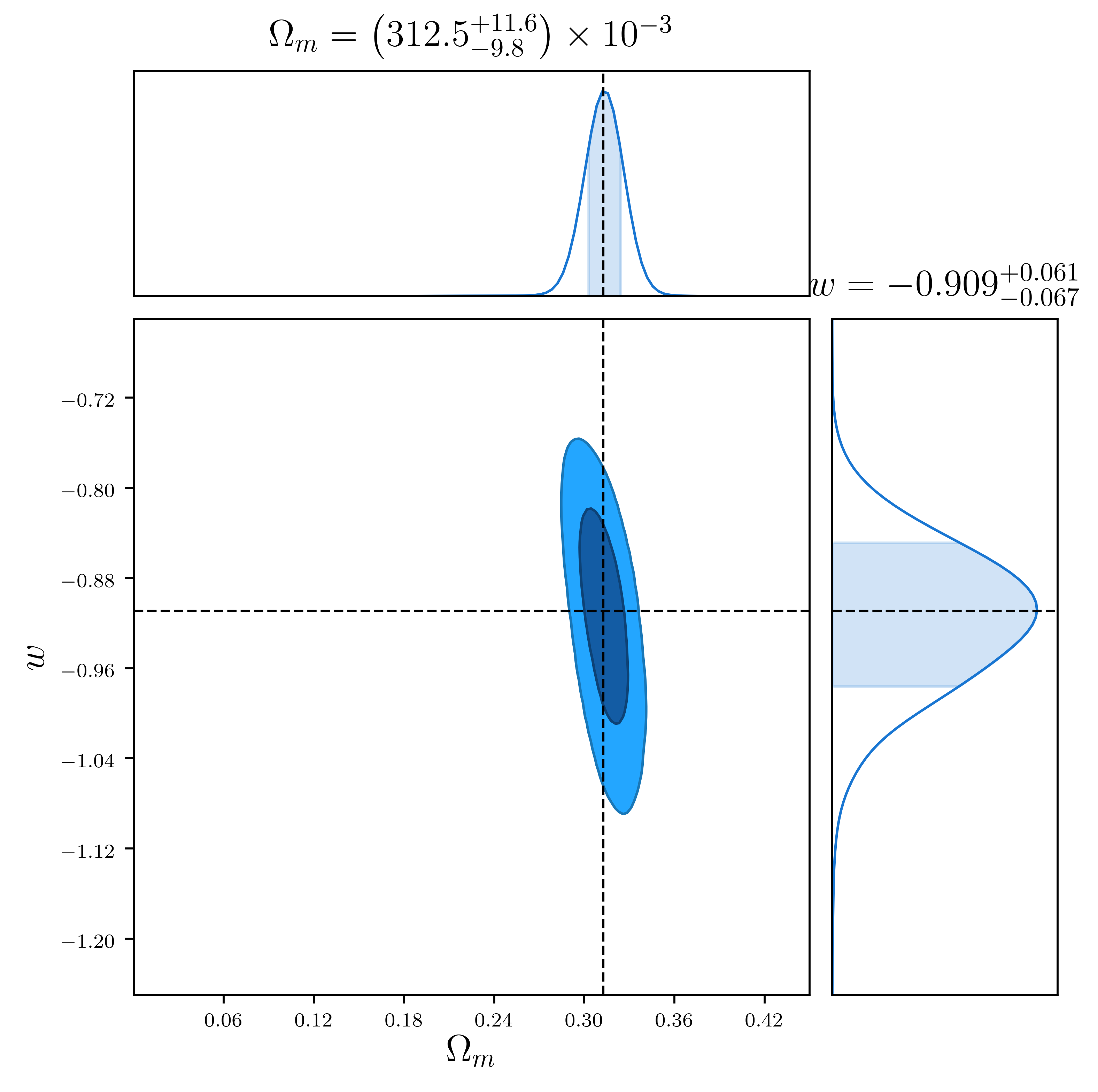}}
\caption{Same as Figure \ref{fig:30}, but with 1001 predicted ages.}
\label{fig:1001}
\end{figure*}

\begin{figure*}[!ht]
{\includegraphics[width=0.3\textwidth]{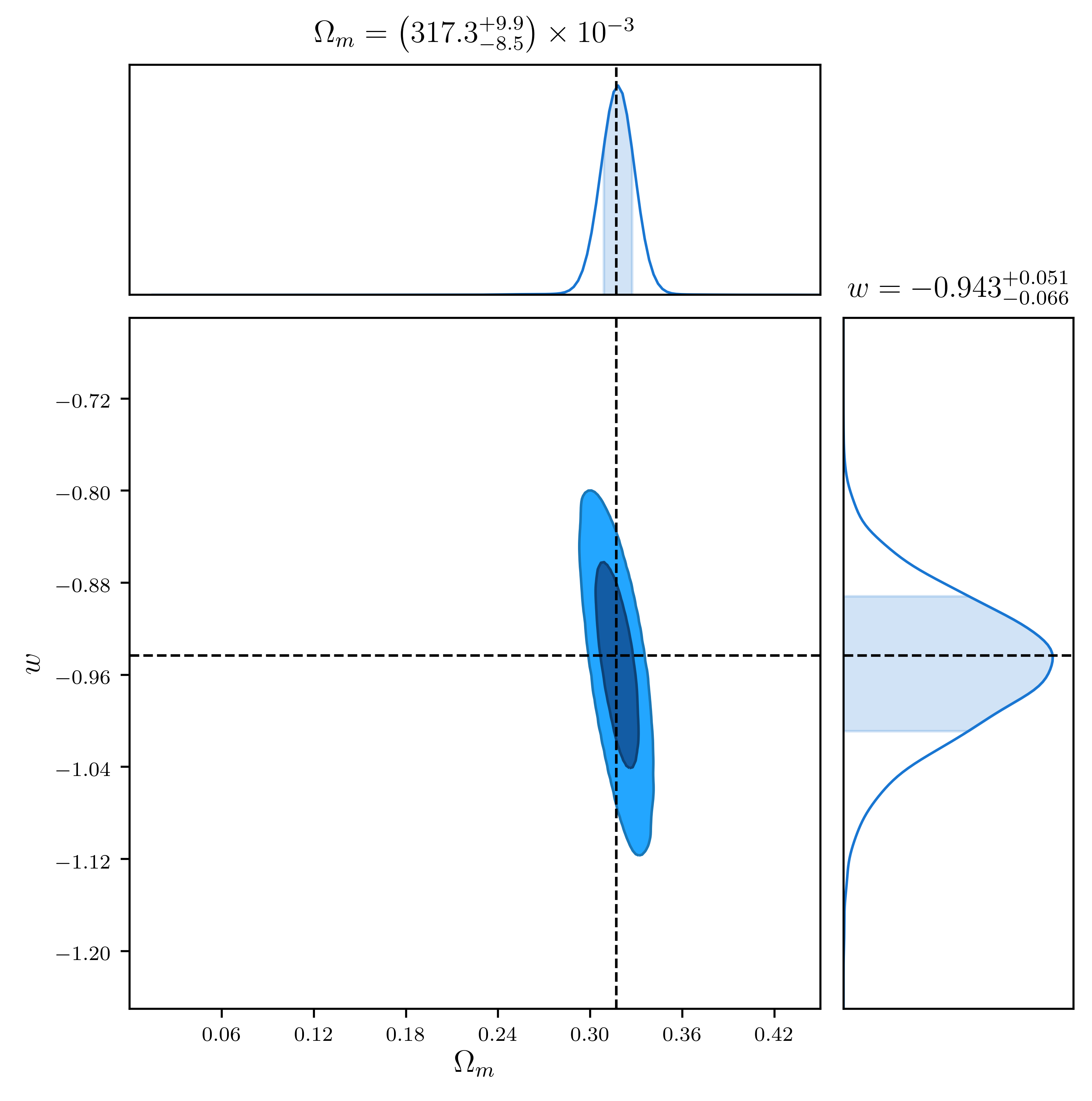}}
{\includegraphics[width=0.3\textwidth]{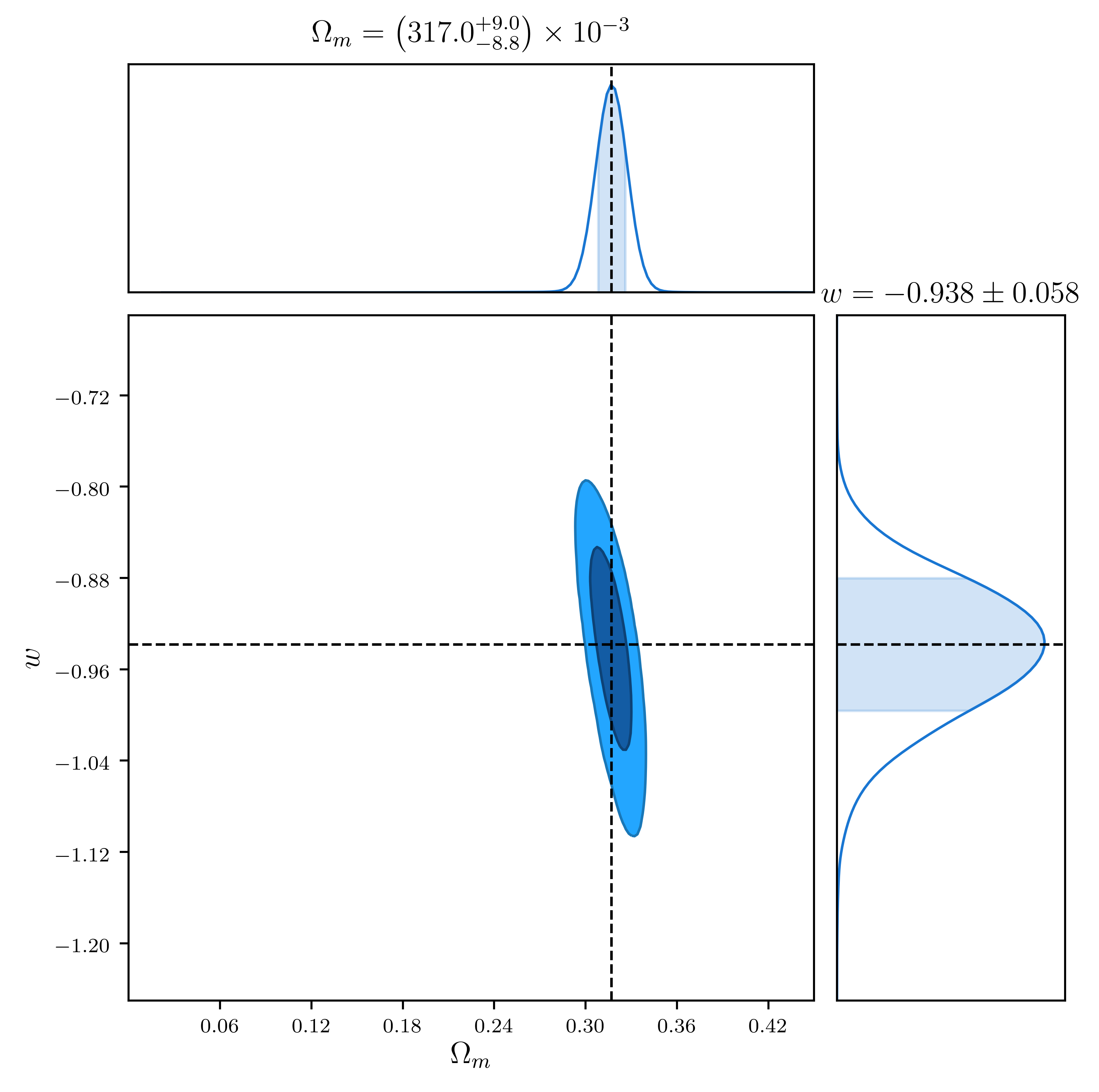}}
{\includegraphics[width=0.3\textwidth]{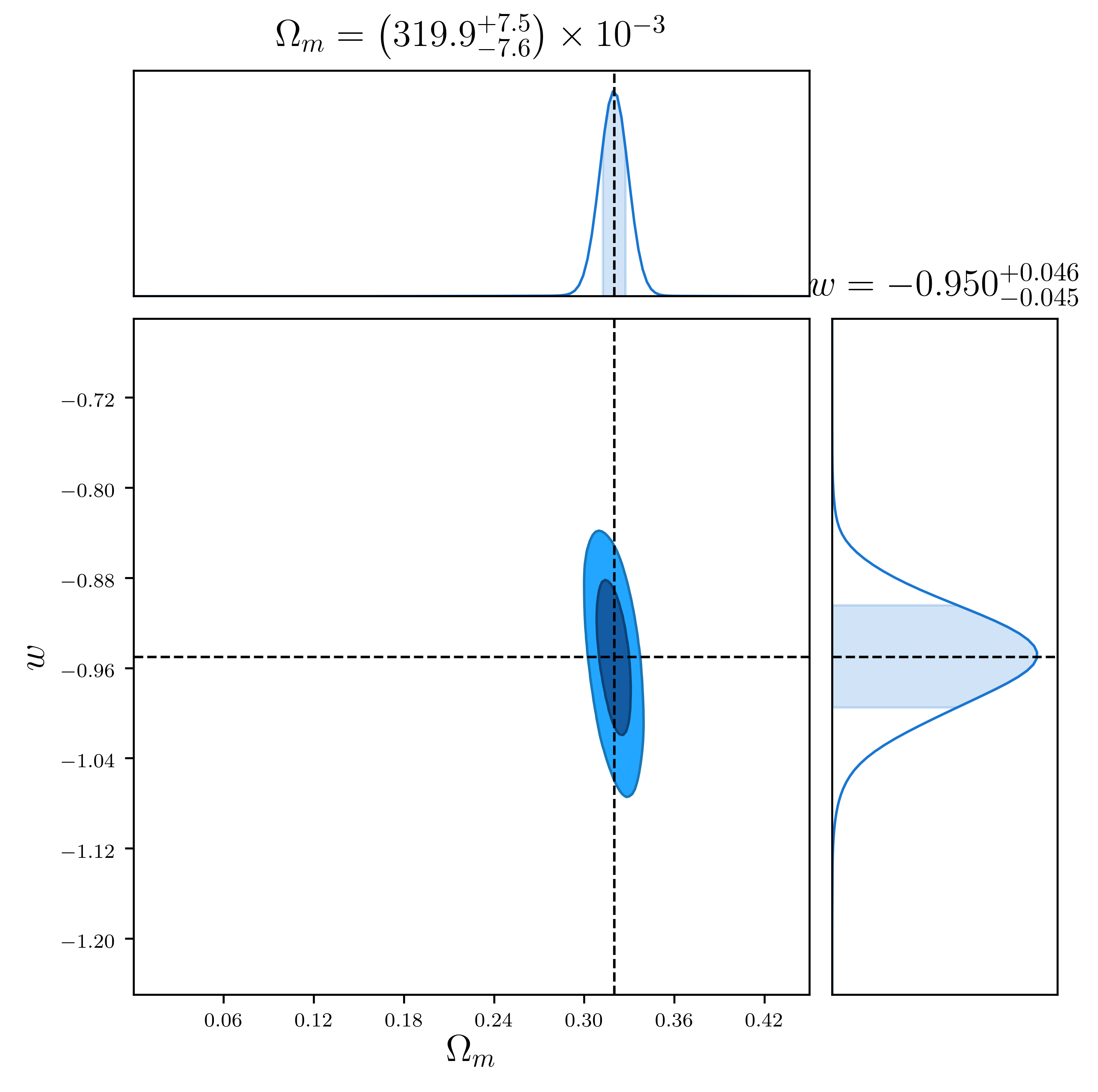}}
\caption{Same as Figure \ref{fig:30}, but with 2004 predicted ages.}

\label{fig:2004}
\end{figure*}

Figures \ref{fig:30}-\ref{fig:2004} show some of the main results of our analyses. In these figures, we show $2\sigma$ contour regions resulting from the analysis involving the age-$z$ relation for the five predicted galaxy age samples using the three ML techniques applied in the $\Omega_m - \omega$ plane. The dashed lines represent the best fits of each parameter, which are illustrated with their respective probability distributions. 

In Figure \ref{fig:30}, we have the parameter space results ($\Omega_m - \omega$) for the predicted sample with 30 points. Note that $\omega$ is quite degenerate in all techniques. Nevertheless, we could obtain the best fits for $\Omega_m$ and $\omega$ in all of them. The SVR technique was the one that best constrained $\omega$. However, the best-fit value found is still distant from what is predicted by the $\Lambda$CDM model. The errors found for the best fits in these analyses, for this sample, are large. Among them, CART presents the worst, as a consequence of having the largest reconstruction error.

However, for SVR, despite finding a best-fit value far from the current cosmology ($\Omega_m = 0.107 \pm{}^{0.084}_{0.105}$ and $\omega = -0.26\pm^{0.049}_{0.136}$), this result is consistent, within the $1\sigma$ error margin, with the values obtained by the authors in \cite{dantas2009}, who conducted an analysis involving a small number of galaxy age points ($\Omega_m \simeq 0.078$ and $\omega= -0.46$). It is necessary to emphasize that in the cited work, marginalization was performed on the $ \tau$ parameter. For a more detailed review of this analysis involving this parameter, see \cite{capozziello2004constraining, pires2006lookback, dantas2007age, dantas2009, dantas2011time}.
Regarding Figures \ref{fig:300}-\ref{fig:1001}, we observe that as we increase the number of sample points, there is a reduction in the degeneracy of the parameter space and in the errors of the best fits. We highlight the sample with 600 predicted points, which presented more consistent best-fits with the observations obtained by \cite{aghanim2020planck}, considering $\Omega_m=0.315\pm{0.007}$ and $\omega=-1.03\pm{0.03}$. For this sample, we also found that the current deceleration parameter of the Universe, $q_0$, which best corresponds to the $\Lambda$CDM scenario, is from the SVR technique, with an approximate value of $q_0 = -0.561\pm {}^{0.101}_{0.126}$ (see Table \ref{q0}). Note also that for the sample with 30 points, we have a decelerating Universe, and this result is not consistent with current observations. However, when considering the $1\sigma$ lower bound error, we have an accelerating Universe for all techniques. ML techniques require a larger number of points for the predictions to agree with the observations, reaffirming what was discussed in previous chapters.
\begin{table}[t]
    \centering
\begin{tabular}{|c|c|c|c|} 
\hline
 Sample &      CART &      MLPR &       SVR \\
 \hline 
 30 &  0.197  $\pm{}^{0.144}_{0.749}$ & 0.057 $\pm{}^{0.213}_{0.551}$ &  0.145$\pm {}^{0.093}_{0.245}$ \\
 300 & -0.419 $\pm {}^{0.164}_{0.218}$ & -0.413 $\pm {}^{0.184}_{0.199}$ & -0.436$\pm {}^{0.112}_{0.148}$ \\
 600 & -0.521 $\pm {}^{0.121}_{0.176}$ & -0.567 $\pm {}^{0.034}_{0.037}$ & -0.562$\pm {}^{0.102}_{0.125}$ \\
1001 & -0.478 $\pm {}^{0.120}_{0.094}$ & -0.442 $\pm {}^{0.097}_{0.100}$ & -0.437$\pm {}^{0.078}_{0.083}$ \\
2004 & -0.466 $\pm {}^{0.065}_{0.080}$ & -0.461 $\pm {}^{0.022}_{0.022}$ & -0.469$\pm {}^{0.036}_{0.083}$ \\
\hline
\end{tabular}
\caption{Values of $q_0$ obtained for the three techniques adopted in this work with their respective predicted samples.}
\label{q0}
\end{table}

As we did previously, the graphs in Figures \ref{fig:2004} show the constraints from 2004 predicted galaxy ages on the parameters $\Omega_m$ and $\omega$. Note that for this sample, we obtained the best parameter space for all techniques, with SVR being the best among them. The best-fit parameters and their respective $1\sigma$ errors for the entire analysis performed in this work, along with the corresponding samples and techniques used, are summarized in Table \ref{inferno}. 

In Table \ref{t0}, we present the current Universe ages obtained using each ML technique and for each simulated sample. These values are expressed in billions of years and are consistent with the age determined by \cite{aghanim2020planck}, which is $13.797 \pm 0.023$ billion years\footnote{This value is derived from an analysis combining TT, TE, EE+lowE+lensing data at $1\sigma$.}, as well as with the upper limit of $t_0 = 13.92{}_{-0.10}^{+0.13}(\text{stat})\pm0.23({\text{sys}})$ billion years obtained from globular cluster studies by \cite{valcin2025ageuniverseglobularclusters}.
\begin{table}[!ht]
\centering

\begin{tabular}{|l|c|c|c|}
 \hline
  Technique & Sample &  $\Omega_m$ & $\omega$ \\
  \hline
   &30 &   0.388$\pm{}^{0.082}_{0.185}$ & -0.330 $\pm{}^{0.130}_{ 0.550}$\\
 &300 &   0.319$\pm{}^{ 0.018}_{0.021}$ & -0.900$\pm{}^{ 0.140}_{0.180}$ \\
   CART&600 &   0.326$\pm{}^{0.013}_{0.014}$ & -1.010 $\pm{}^{0.100}_{0.150}$ \\
  &1001 &   0.320 $\pm{}^{0.011}_{0.012}$& -0.959 $\pm{}^{0.087}_{ 0.090}$ \\
   & 2004 &   0.317 $\pm{}^{0.010}_{0.010}$& -0.943$\pm{}^{0.051}_{0.066}$  \\
   \hline
  &30 &   0.314 $\pm{}^{0.074}_{0.112}$& -0.430$\pm{}^{0.180}_{0.400}$  \\
   &300 &   0.316 $\pm{}^{0.018}_{ 0.022}$& -0.890$\pm{}^{0.160}_{0.160}$  \\
  MLPR &600 &   0.329$\pm{}^{ 0.014}_{0.014}$ & -1.060 $\pm{}^{0.120}_{ 0.140}$ \\
   &1001 &   0.315$\pm{}^{0.012}_{0.011}$ & -0.917$\pm{}^{0.080}_{0.081}$ \\
   &2004 &   0.317$\pm{}^{0.009}_{0.009}$ & -0.938 $\pm{}^{0.058}_{0.058}$ \\
   \hline
  &  30 &   0.107 $\pm{}^{0.084}_{0.105}$& -0.265$\pm{}^{0.049}_{0.136}$ \\
  & 300 &   0.316$\pm{}^{0.013}_{0.012}$ & -0.912$\pm{}^{0.094}_{0.126}$ \\
  SVR & 600 &   0.329$\pm{}^{0.010}_{0.010}$ & -1.054 $\pm{}^{0.087}_{0.107}$\\
  & 1001 &   0.313$\pm{}^{ 0.012}_{0.010}$ & -0.909$\pm{}^{0.061}_{0.067}$\\
   & 2004 &   0.320$\pm{}^{0.010}_{0.010}$ & -0.950 $\pm{}^{0.046}_{0.045}$ \\
   \hline
\end{tabular}
    
    \caption{Results of the best-fit values of $\Omega_m$ and $\omega$ for the three techniques.}
    \label{inferno}

\end{table}

\begin{table}[!ht]
    \centering
\begin{tabular}{|c|c|c|c|}
\hline
 Sample & CART & MLPR & SVR \\
 \hline
30 &  $11.399\pm{}^{0.758}_{3.817}$ &  $12.183\pm{}^{1.122}_{2.86}$ &  $12.340\pm{}^{0.839}_{3.679}$  \\
300 & $13.616\pm{}^{0.521}_{0.647}$  &  $13.628\pm{}^{0.581}_{0.629}$  &  $13.678\pm{}^{0.361}_{0.419}$  \\
600 &  $13.764\pm{}^{0.668}_{0.454}$  & $13.821\pm{}^{0.774}_{0.202}$   &  $13.816\pm{}^{0.227}_{0.314}$  \\
1001 & $13.733\pm{}^{0.312}_{0.333}$  & $13.700\pm{}^{0.318}_{0.312}$  & $13.712\pm{}^{0.272}_{0.270}$  \\
2004 & $13.731\pm{}^{0.224}_{0.244}$  & $13.723\pm{}^{0.230}_{0.231}$  & $13.715\pm{}^{0.185}_{0.187}$ 
\\
\hline
\end{tabular}
    \caption{Current Universe ages obtained for each ML technique and their respective samples. These ages are presented in billions of years.}
\label{t0}
\end{table}

In order to corroborate the methodology adopted in our work, we also calculated the best fit for the simulated samples (without using ML) and found, for example, for the samples with 2004 points, values of the order of $\Omega_m= 0.323\pm^{0.007}_{0.007}$, $\omega = -1.01\pm^{0.051}_{0.052}$. Note that the best fits for this sample are consistent with all the best fits of the sample with 600 predicted points for all three techniques (see Table \ref{inferno}). This indicates that the predictions of our samples represent the base dataset very well. Thus, by using 600 predicted points to obtain the contour regions, we achieved best-fits equivalent to a sample of 2004 simulated points with a shorter computational time.
Another factor influencing computational performance is the number of hyperparameters used, especially in the MLPR technique, which requires a greater number to be adjusted. For this reason, the number of hyperparameters used during optimization was smaller, which was a determining factor in the errors found for this technique being higher than those found via SVR.
Techniques like CART require more extensive treatment to obtain an appropriate tree. Additionally, they are unstable because small variations in the data result in the generation of a completely different tree. In our case, the max\_depth directly influences the value of the predicted age, making it necessary to pay extra attention to this hyperparameter.
\section{Conclusion}\label{conclusion}
ML methods have proven effective in detecting patterns in data without human intervention and using these patterns to make predictions or decisions under uncertainty. In this study, we employed supervised learning techniques — CART, MLPR, and SVR — to reconstruct galaxy ages using simulated data samples of varying sizes (100, 1000, 2000, 3334, 6680) generated by the Monte Carlo Method (MCM). The results demonstrated that the SVR technique performed the best, accurately capturing the nonlinear behavior of the data while avoiding overfitting. The CART method, despite reproducing nonlinear behavior, yielded higher MSEs and less satisfactory curves due to the piecewise constant nature of decision tree predictions.

We achieved the best fits for the cosmological parameters $\Omega_m$ and $\omega$ in our adopted flat $\Lambda$CDM model. Specifically, the 600-point predicted sample provided the most consistent best fits with \citep{aghanim2020planck}, with the SVR technique yielding a deceleration parameter $q_0 = -0.561 \pm^{0.101}_{0.126}$. The reconstructed galaxy ages are consistent with the Planck Collaboration's estimate of $13.797 \pm 0.023$ billion years.

We found that the errors at $1\sigma$ were significant and varied depending on the sample size and technique. The 2004-point sample exhibited the lowest error, with the SVR predicting an age of $t_0 = 13.715 \pm^{0.185}_{0.187}$ billion years. A complementary analysis of the 2004-point simulated sample (without ML) produced cosmological parameter values consistent with the best fits from the 600-point predicted sample.

In conclusion, the predicted data effectively reproduces the base dataset for reconstruction. Using predicted data to calculate free parameters resulted in best fits comparable to those obtained from the entire sample, with reduced computational time. The use of ML techniques, especially SVR - which achieved the best performance in our analysis - for reconstructing galaxy ages from observational data holds great potential for cosmology, as it can improve predictive accuracy and help place tighter constraints on key cosmological parameters.

\begin{acknowledgments}
ASN thanks the Coordenação de Aperfeiçoamento de Pessoal de Nível Superior (CAPES) for the grant under which this work was carried out. ASN also thanks Camila Franco and Carlos Bengaly for the valuable discussions on cosmology and their helpful comments on the manuscript. MAD is grateful to the Universidade do Estado do Rio Grande do Norte (UERN).
\end{acknowledgments}

\appendix
\bibliography{bib}

\begin{thebibliography}{39}%
\makeatletter
\providecommand \@ifxundefined [1]{%
 \@ifx{#1\undefined}
}%
\providecommand \@ifnum [1]{%
 \ifnum #1\expandafter \@firstoftwo
 \else \expandafter \@secondoftwo
 \fi
}%
\providecommand \@ifx [1]{%
 \ifx #1\expandafter \@firstoftwo
 \else \expandafter \@secondoftwo
 \fi
}%
\providecommand \natexlab [1]{#1}%
\providecommand \enquote  [1]{``#1''}%
\providecommand \bibnamefont  [1]{#1}%
\providecommand \bibfnamefont [1]{#1}%
\providecommand \citenamefont [1]{#1}%
\providecommand \href@noop [0]{\@secondoftwo}%
\providecommand \href [0]{\begingroup \@sanitize@url \@href}%
\providecommand \@href[1]{\@@startlink{#1}\@@href}%
\providecommand \@@href[1]{\endgroup#1\@@endlink}%
\providecommand \@sanitize@url [0]{\catcode `\\12\catcode `\$12\catcode `\&12\catcode `\#12\catcode `\^12\catcode `\_12\catcode `\%12\relax}%
\providecommand \@@startlink[1]{}%
\providecommand \@@endlink[0]{}%
\providecommand \url  [0]{\begingroup\@sanitize@url \@url }%
\providecommand \@url [1]{\endgroup\@href {#1}{\urlprefix }}%
\providecommand \urlprefix  [0]{URL }%
\providecommand \Eprint [0]{\href }%
\providecommand \doibase [0]{https://doi.org/}%
\providecommand \selectlanguage [0]{\@gobble}%
\providecommand \bibinfo  [0]{\@secondoftwo}%
\providecommand \bibfield  [0]{\@secondoftwo}%
\providecommand \translation [1]{[#1]}%
\providecommand \BibitemOpen [0]{}%
\providecommand \bibitemStop [0]{}%
\providecommand \bibitemNoStop [0]{.\EOS\space}%
\providecommand \EOS [0]{\spacefactor3000\relax}%
\providecommand \BibitemShut  [1]{\csname bibitem#1\endcsname}%
\let\auto@bib@innerbib\@empty
\bibitem [{\citenamefont {Ostriker}\ and\ \citenamefont {Steinhardt}(1995)}]{ostriker1995cosmicconcordance}%
  \BibitemOpen
  \bibfield  {author} {\bibinfo {author} {\bibfnamefont {J.~P.}\ \bibnamefont {Ostriker}}\ and\ \bibinfo {author} {\bibfnamefont {P.~J.}\ \bibnamefont {Steinhardt}},\ }\href {https://arxiv.org/abs/astro-ph/9505066} {\bibinfo {title} {Cosmic concordance}} (\bibinfo {year} {1995}),\ \Eprint {https://arxiv.org/abs/astro-ph/9505066} {arXiv:astro-ph/9505066 [astro-ph]} \BibitemShut {NoStop}%
\bibitem [{\citenamefont {Jimenez}(1997)}]{jimenez1997ageuniverse}%
  \BibitemOpen
  \bibfield  {author} {\bibinfo {author} {\bibfnamefont {R.}~\bibnamefont {Jimenez}},\ }\href {https://arxiv.org/abs/astro-ph/9701222} {\bibinfo {title} {The age of the universe}} (\bibinfo {year} {1997}),\ \Eprint {https://arxiv.org/abs/astro-ph/9701222} {arXiv:astro-ph/9701222 [astro-ph]} \BibitemShut {NoStop}%
\bibitem [{\citenamefont {Chaboyer}(1998)}]{Chaboyer_1998}%
  \BibitemOpen
  \bibfield  {author} {\bibinfo {author} {\bibfnamefont {B.}~\bibnamefont {Chaboyer}},\ }\href {https://doi.org/10.1016/s0370-1573(98)00054-4} {\bibfield  {journal} {\bibinfo  {journal} {Physics Reports}\ }\textbf {\bibinfo {volume} {307}},\ \bibinfo {pages} {23–30} (\bibinfo {year} {1998})}\BibitemShut {NoStop}%
\bibitem [{\citenamefont {Krauss}\ and\ \citenamefont {Turner}(1995)}]{Krauss_1995}%
  \BibitemOpen
  \bibfield  {author} {\bibinfo {author} {\bibfnamefont {L.~M.}\ \bibnamefont {Krauss}}\ and\ \bibinfo {author} {\bibfnamefont {M.~S.}\ \bibnamefont {Turner}},\ }\href {https://doi.org/10.1007/bf02108229} {\bibfield  {journal} {\bibinfo  {journal} {General Relativity and Gravitation}\ }\textbf {\bibinfo {volume} {27}},\ \bibinfo {pages} {1137–1144} (\bibinfo {year} {1995})}\BibitemShut {NoStop}%
\bibitem [{\citenamefont {{Krauss}}\ and\ \citenamefont {{Chaboyer}}(2003)}]{2003Sci...299...65K}%
  \BibitemOpen
  \bibfield  {author} {\bibinfo {author} {\bibfnamefont {L.~M.}\ \bibnamefont {{Krauss}}}\ and\ \bibinfo {author} {\bibfnamefont {B.}~\bibnamefont {{Chaboyer}}},\ }\href {https://doi.org/10.1126/science.1075631} {\bibfield  {journal} {\bibinfo  {journal} {Science}\ }\textbf {\bibinfo {volume} {299}},\ \bibinfo {pages} {65} (\bibinfo {year} {2003})}\BibitemShut {NoStop}%
\bibitem [{\citenamefont {Jimenez}\ and\ \citenamefont {Loeb}(2002)}]{Jimenez_2002}%
  \BibitemOpen
  \bibfield  {author} {\bibinfo {author} {\bibfnamefont {R.}~\bibnamefont {Jimenez}}\ and\ \bibinfo {author} {\bibfnamefont {A.}~\bibnamefont {Loeb}},\ }\href {https://doi.org/10.1086/340549} {\bibfield  {journal} {\bibinfo  {journal} {The Astrophysical Journal}\ }\textbf {\bibinfo {volume} {573}},\ \bibinfo {pages} {37–42} (\bibinfo {year} {2002})}\BibitemShut {NoStop}%
\bibitem [{\citenamefont {Simon}\ \emph {et~al.}(2005)\citenamefont {Simon}, \citenamefont {Verde},\ and\ \citenamefont {Jimenez}}]{simon2005constraints}%
  \BibitemOpen
  \bibfield  {author} {\bibinfo {author} {\bibfnamefont {J.}~\bibnamefont {Simon}}, \bibinfo {author} {\bibfnamefont {L.}~\bibnamefont {Verde}},\ and\ \bibinfo {author} {\bibfnamefont {R.}~\bibnamefont {Jimenez}},\ }\href@noop {} {\bibfield  {journal} {\bibinfo  {journal} {Physical Review D}\ }\textbf {\bibinfo {volume} {71}},\ \bibinfo {pages} {123001} (\bibinfo {year} {2005})}\BibitemShut {NoStop}%
\bibitem [{\citenamefont {Nolan}\ \emph {et~al.}(2001)\citenamefont {Nolan}, \citenamefont {Dunlop},\ and\ \citenamefont {Jimenez}}]{nolan2001sun}%
  \BibitemOpen
  \bibfield  {author} {\bibinfo {author} {\bibfnamefont {L.~A.}\ \bibnamefont {Nolan}}, \bibinfo {author} {\bibfnamefont {J.~S.}\ \bibnamefont {Dunlop}},\ and\ \bibinfo {author} {\bibfnamefont {R.}~\bibnamefont {Jimenez}},\ }\href@noop {} {\bibfield  {journal} {\bibinfo  {journal} {Monthly Notices of the Royal Astronomical Society}\ }\textbf {\bibinfo {volume} {323}},\ \bibinfo {pages} {385} (\bibinfo {year} {2001})}\BibitemShut {NoStop}%
\bibitem [{\citenamefont {Vagnozzi}\ \emph {et~al.}(2022)\citenamefont {Vagnozzi}, \citenamefont {Pacucci},\ and\ \citenamefont {Loeb}}]{Vagnozzi_2022}%
  \BibitemOpen
  \bibfield  {author} {\bibinfo {author} {\bibfnamefont {S.}~\bibnamefont {Vagnozzi}}, \bibinfo {author} {\bibfnamefont {F.}~\bibnamefont {Pacucci}},\ and\ \bibinfo {author} {\bibfnamefont {A.}~\bibnamefont {Loeb}},\ }\href {https://doi.org/10.1016/j.jheap.2022.07.004} {\bibfield  {journal} {\bibinfo  {journal} {Journal of High Energy Astrophysics}\ }\textbf {\bibinfo {volume} {36}},\ \bibinfo {pages} {27–35} (\bibinfo {year} {2022})}\BibitemShut {NoStop}%
\bibitem [{\citenamefont {Wei}\ and\ \citenamefont {Melia}(2022)}]{Wei_2022}%
  \BibitemOpen
  \bibfield  {author} {\bibinfo {author} {\bibfnamefont {J.-J.}\ \bibnamefont {Wei}}\ and\ \bibinfo {author} {\bibfnamefont {F.}~\bibnamefont {Melia}},\ }\href {https://doi.org/10.3847/1538-4357/ac562c} {\bibfield  {journal} {\bibinfo  {journal} {The Astrophysical Journal}\ }\textbf {\bibinfo {volume} {928}},\ \bibinfo {pages} {165} (\bibinfo {year} {2022})}\BibitemShut {NoStop}%
\bibitem [{\citenamefont {Dantas}\ \emph {et~al.}(2007)\citenamefont {Dantas}, \citenamefont {Alcaniz}, \citenamefont {Jain},\ and\ \citenamefont {Dev}}]{dantas2007age}%
  \BibitemOpen
  \bibfield  {author} {\bibinfo {author} {\bibfnamefont {M.}~\bibnamefont {Dantas}}, \bibinfo {author} {\bibfnamefont {J.}~\bibnamefont {Alcaniz}}, \bibinfo {author} {\bibfnamefont {D.}~\bibnamefont {Jain}},\ and\ \bibinfo {author} {\bibfnamefont {A.}~\bibnamefont {Dev}},\ }\href@noop {} {\bibfield  {journal} {\bibinfo  {journal} {Astronomy \& Astrophysics}\ }\textbf {\bibinfo {volume} {467}},\ \bibinfo {pages} {421} (\bibinfo {year} {2007})}\BibitemShut {NoStop}%
\bibitem [{\citenamefont {Dantas}\ \emph {et~al.}(2011)\citenamefont {Dantas}, \citenamefont {Alcaniz}, \citenamefont {Mania},\ and\ \citenamefont {Ratra}}]{dantas2011time}%
  \BibitemOpen
  \bibfield  {author} {\bibinfo {author} {\bibfnamefont {M.}~\bibnamefont {Dantas}}, \bibinfo {author} {\bibfnamefont {J.}~\bibnamefont {Alcaniz}}, \bibinfo {author} {\bibfnamefont {D.}~\bibnamefont {Mania}},\ and\ \bibinfo {author} {\bibfnamefont {B.}~\bibnamefont {Ratra}},\ }\href@noop {} {\bibfield  {journal} {\bibinfo  {journal} {Physics Letters B}\ }\textbf {\bibinfo {volume} {699}},\ \bibinfo {pages} {239} (\bibinfo {year} {2011})}\BibitemShut {NoStop}%
\bibitem [{\citenamefont {Dantas}\ \emph {et~al.}(2009)\citenamefont {Dantas}, \citenamefont {Alcaniz},\ and\ \citenamefont {Pires}}]{dantas2009}%
  \BibitemOpen
  \bibfield  {author} {\bibinfo {author} {\bibfnamefont {M.}~\bibnamefont {Dantas}}, \bibinfo {author} {\bibfnamefont {J.}~\bibnamefont {Alcaniz}},\ and\ \bibinfo {author} {\bibfnamefont {N.}~\bibnamefont {Pires}},\ }\href {https://doi.org/10.1016/j.physletb.2009.08.008} {\bibfield  {journal} {\bibinfo  {journal} {Physics Letters B}\ }\textbf {\bibinfo {volume} {679}},\ \bibinfo {pages} {423–427} (\bibinfo {year} {2009})}\BibitemShut {NoStop}%
\bibitem [{\citenamefont {Capozziello}\ \emph {et~al.}(2004)\citenamefont {Capozziello}, \citenamefont {Cardone}, \citenamefont {Funaro},\ and\ \citenamefont {Andreon}}]{capozziello2004constraining}%
  \BibitemOpen
  \bibfield  {author} {\bibinfo {author} {\bibfnamefont {S.}~\bibnamefont {Capozziello}}, \bibinfo {author} {\bibfnamefont {V.~F.}\ \bibnamefont {Cardone}}, \bibinfo {author} {\bibfnamefont {M.}~\bibnamefont {Funaro}},\ and\ \bibinfo {author} {\bibfnamefont {S.}~\bibnamefont {Andreon}},\ }\href@noop {} {\bibfield  {journal} {\bibinfo  {journal} {Physical Review D}\ }\textbf {\bibinfo {volume} {70}},\ \bibinfo {pages} {123501} (\bibinfo {year} {2004})}\BibitemShut {NoStop}%
\bibitem [{\citenamefont {Pires}\ \emph {et~al.}(2006)\citenamefont {Pires}, \citenamefont {Zhu},\ and\ \citenamefont {Alcaniz}}]{pires2006lookback}%
  \BibitemOpen
  \bibfield  {author} {\bibinfo {author} {\bibfnamefont {N.}~\bibnamefont {Pires}}, \bibinfo {author} {\bibfnamefont {Z.-H.}\ \bibnamefont {Zhu}},\ and\ \bibinfo {author} {\bibfnamefont {J.~S.}\ \bibnamefont {Alcaniz}},\ }\href@noop {} {\bibfield  {journal} {\bibinfo  {journal} {Physical Review D}\ }\textbf {\bibinfo {volume} {73}},\ \bibinfo {pages} {123530} (\bibinfo {year} {2006})}\BibitemShut {NoStop}%
\bibitem [{\citenamefont {Binici}\ \emph {et~al.}(2024)\citenamefont {Binici}, \citenamefont {Deliduman},\ and\ \citenamefont {Şakir Dilsiz}}]{binici2024age}%
  \BibitemOpen
  \bibfield  {author} {\bibinfo {author} {\bibfnamefont {S.~S.}\ \bibnamefont {Binici}}, \bibinfo {author} {\bibfnamefont {C.}~\bibnamefont {Deliduman}},\ and\ \bibinfo {author} {\bibfnamefont {F.}~\bibnamefont {Şakir Dilsiz}},\ }\href {https://arxiv.org/abs/2402.16646} {\bibinfo {title} {The ages of the oldest astrophysical objects in an ellipsoidal universe}} (\bibinfo {year} {2024}),\ \Eprint {https://arxiv.org/abs/2402.16646} {arXiv:2402.16646 [astro-ph.CO]} \BibitemShut {NoStop}%
\bibitem [{\citenamefont {Gao}\ \emph {et~al.}(2024)\citenamefont {Gao}, \citenamefont {López-Corredoira},\ and\ \citenamefont {Wei}}]{Gao_2024}%
  \BibitemOpen
  \bibfield  {author} {\bibinfo {author} {\bibfnamefont {C.-Y.}\ \bibnamefont {Gao}}, \bibinfo {author} {\bibfnamefont {M.}~\bibnamefont {López-Corredoira}},\ and\ \bibinfo {author} {\bibfnamefont {J.-J.}\ \bibnamefont {Wei}},\ }\href {https://doi.org/10.3847/1538-4357/ad5ce4} {\bibfield  {journal} {\bibinfo  {journal} {The Astrophysical Journal}\ }\textbf {\bibinfo {volume} {970}},\ \bibinfo {pages} {142} (\bibinfo {year} {2024})}\BibitemShut {NoStop}%
\bibitem [{\citenamefont {{Castro-Rodr{\'\i}guez}}\ and\ \citenamefont {{L{\'o}pez-Corredoira}}(2012)}]{2012A&A...537A..31C}%
  \BibitemOpen
  \bibfield  {author} {\bibinfo {author} {\bibfnamefont {N.}~\bibnamefont {{Castro-Rodr{\'\i}guez}}}\ and\ \bibinfo {author} {\bibfnamefont {M.}~\bibnamefont {{L{\'o}pez-Corredoira}}},\ }\href {https://doi.org/10.1051/0004-6361/201117418} {\bibfield  {journal} {\bibinfo  {journal} {\aap}\ }\textbf {\bibinfo {volume} {537}},\ \bibinfo {eid} {A31} (\bibinfo {year} {2012})},\ \Eprint {https://arxiv.org/abs/1111.2726} {arXiv:1111.2726 [astro-ph.CO]} \BibitemShut {NoStop}%
\bibitem [{\citenamefont {Chac{\'o}n}\ \emph {et~al.}(2021)\citenamefont {Chac{\'o}n}, \citenamefont {V{\'a}zquez},\ and\ \citenamefont {Almaraz}}]{chacon2021classification}%
  \BibitemOpen
  \bibfield  {author} {\bibinfo {author} {\bibfnamefont {J.}~\bibnamefont {Chac{\'o}n}}, \bibinfo {author} {\bibfnamefont {J.~A.}\ \bibnamefont {V{\'a}zquez}},\ and\ \bibinfo {author} {\bibfnamefont {E.}~\bibnamefont {Almaraz}},\ }\href@noop {} {\bibfield  {journal} {\bibinfo  {journal} {arXiv preprint arXiv:2106.06587}\ } (\bibinfo {year} {2021})}\BibitemShut {NoStop}%
\bibitem [{\citenamefont {Arjona}\ \emph {et~al.}(2021)\citenamefont {Arjona}, \citenamefont {Melchiorri},\ and\ \citenamefont {Nesseris}}]{arjona2021testing}%
  \BibitemOpen
  \bibfield  {author} {\bibinfo {author} {\bibfnamefont {R.}~\bibnamefont {Arjona}}, \bibinfo {author} {\bibfnamefont {A.}~\bibnamefont {Melchiorri}},\ and\ \bibinfo {author} {\bibfnamefont {S.}~\bibnamefont {Nesseris}},\ }\href@noop {} {\bibinfo {title} {Testing the $\lambda$cdm paradigm with growth rate data and machine learning}} (\bibinfo {year} {2021}),\ \Eprint {https://arxiv.org/abs/2107.04343} {arXiv:2107.04343 [astro-ph.CO]} \BibitemShut {NoStop}%
\bibitem [{\citenamefont {von Marttens}\ \emph {et~al.}(2021)\citenamefont {von Marttens}, \citenamefont {Casarini}, \citenamefont {Napolitano}, \citenamefont {Wu}, \citenamefont {Amaro}, \citenamefont {Li}, \citenamefont {Tortora}, \citenamefont {Canabarro},\ and\ \citenamefont {Wang}}]{von2021inferring}%
  \BibitemOpen
  \bibfield  {author} {\bibinfo {author} {\bibfnamefont {R.}~\bibnamefont {von Marttens}}, \bibinfo {author} {\bibfnamefont {L.}~\bibnamefont {Casarini}}, \bibinfo {author} {\bibfnamefont {N.~R.}\ \bibnamefont {Napolitano}}, \bibinfo {author} {\bibfnamefont {S.}~\bibnamefont {Wu}}, \bibinfo {author} {\bibfnamefont {V.}~\bibnamefont {Amaro}}, \bibinfo {author} {\bibfnamefont {R.}~\bibnamefont {Li}}, \bibinfo {author} {\bibfnamefont {C.}~\bibnamefont {Tortora}}, \bibinfo {author} {\bibfnamefont {A.}~\bibnamefont {Canabarro}},\ and\ \bibinfo {author} {\bibfnamefont {Y.}~\bibnamefont {Wang}},\ }\href@noop {} {\bibfield  {journal} {\bibinfo  {journal} {arXiv preprint arXiv:2111.01185}\ } (\bibinfo {year} {2021})}\BibitemShut {NoStop}%
\bibitem [{\citenamefont {{Bengaly}}\ \emph {et~al.}(2023)\citenamefont {{Bengaly}}, \citenamefont {{Dantas}}, \citenamefont {{Casarini}},\ and\ \citenamefont {{Alcaniz}}}]{carlos2023}%
  \BibitemOpen
  \bibfield  {author} {\bibinfo {author} {\bibfnamefont {C.}~\bibnamefont {{Bengaly}}}, \bibinfo {author} {\bibfnamefont {M.~A.}\ \bibnamefont {{Dantas}}}, \bibinfo {author} {\bibfnamefont {L.}~\bibnamefont {{Casarini}}},\ and\ \bibinfo {author} {\bibfnamefont {J.}~\bibnamefont {{Alcaniz}}},\ }\href {https://doi.org/10.1140/epjc/s10052-023-11734-1} {\bibfield  {journal} {\bibinfo  {journal} {European Physical Journal C}\ }\textbf {\bibinfo {volume} {83}},\ \bibinfo {eid} {548} (\bibinfo {year} {2023})},\ \Eprint {https://arxiv.org/abs/2209.09017} {arXiv:2209.09017 [astro-ph.CO]} \BibitemShut {NoStop}%
\bibitem [{\citenamefont {{Pal}}\ and\ \citenamefont {{Saha}}(2024)}]{2024PhyS...99k5007P}%
  \BibitemOpen
  \bibfield  {author} {\bibinfo {author} {\bibfnamefont {S.}~\bibnamefont {{Pal}}}\ and\ \bibinfo {author} {\bibfnamefont {R.}~\bibnamefont {{Saha}}},\ }\href {https://doi.org/10.1088/1402-4896/ad804d} {\bibfield  {journal} {\bibinfo  {journal} {\physscr}\ }\textbf {\bibinfo {volume} {99}},\ \bibinfo {eid} {115007} (\bibinfo {year} {2024})},\ \Eprint {https://arxiv.org/abs/2309.15179} {arXiv:2309.15179 [astro-ph.CO]} \BibitemShut {NoStop}%
\bibitem [{\citenamefont {Mukherjee}\ and\ \citenamefont {Sen}(2025)}]{mukherjee2025newsim5sigmatension}%
  \BibitemOpen
  \bibfield  {author} {\bibinfo {author} {\bibfnamefont {P.}~\bibnamefont {Mukherjee}}\ and\ \bibinfo {author} {\bibfnamefont {A.~A.}\ \bibnamefont {Sen}},\ }\href {https://arxiv.org/abs/2503.02880} {\bibinfo {title} {A new $\sim 5\sigma$ tension at characteristic redshift from desi-dr1 bao and des-sn5yr observations}} (\bibinfo {year} {2025}),\ \Eprint {https://arxiv.org/abs/2503.02880} {arXiv:2503.02880 [astro-ph.CO]} \BibitemShut {NoStop}%
\bibitem [{\citenamefont {Borghetto}\ \emph {et~al.}(2025)\citenamefont {Borghetto}, \citenamefont {Malhotra}, \citenamefont {Tasinato},\ and\ \citenamefont {Zavala}}]{borghetto2025boundeddarkenergy}%
  \BibitemOpen
  \bibfield  {author} {\bibinfo {author} {\bibfnamefont {G.}~\bibnamefont {Borghetto}}, \bibinfo {author} {\bibfnamefont {A.}~\bibnamefont {Malhotra}}, \bibinfo {author} {\bibfnamefont {G.}~\bibnamefont {Tasinato}},\ and\ \bibinfo {author} {\bibfnamefont {I.}~\bibnamefont {Zavala}},\ }\href {https://arxiv.org/abs/2503.11628} {\bibinfo {title} {Bounded dark energy}} (\bibinfo {year} {2025}),\ \Eprint {https://arxiv.org/abs/2503.11628} {arXiv:2503.11628 [astro-ph.CO]} \BibitemShut {NoStop}%
\bibitem [{\citenamefont {Metropolis}\ \emph {et~al.}(1987)\citenamefont {Metropolis} \emph {et~al.}}]{metropolis1987beginning}%
  \BibitemOpen
  \bibfield  {author} {\bibinfo {author} {\bibfnamefont {N.}~\bibnamefont {Metropolis}} \emph {et~al.},\ }\href@noop {} {\bibfield  {journal} {\bibinfo  {journal} {Los Alamos Science}\ }\textbf {\bibinfo {volume} {15}},\ \bibinfo {pages} {125} (\bibinfo {year} {1987})}\BibitemShut {NoStop}%
\bibitem [{\citenamefont {{Sandage}}(1988)}]{1988Sandage}%
  \BibitemOpen
  \bibfield  {author} {\bibinfo {author} {\bibfnamefont {A.}~\bibnamefont {{Sandage}}},\ }\href {https://doi.org/10.1146/annurev.aa.26.090188.003021} {\bibfield  {journal} {\bibinfo  {journal} {\araa}\ }\textbf {\bibinfo {volume} {26}},\ \bibinfo {pages} {561} (\bibinfo {year} {1988})}\BibitemShut {NoStop}%
\bibitem [{\citenamefont {Peebles}(1993)}]{1993peebles}%
  \BibitemOpen
  \bibfield  {author} {\bibinfo {author} {\bibfnamefont {P.~J.~E.}\ \bibnamefont {Peebles}},\ }\href@noop {} {\emph {\bibinfo {title} {Principles of physical cosmology}}},\ Vol.~\bibinfo {volume} {27}\ (\bibinfo  {publisher} {Princeton university press},\ \bibinfo {year} {1993})\BibitemShut {NoStop}%
\bibitem [{\citenamefont {Savage}\ \emph {et~al.}(2005)\citenamefont {Savage}, \citenamefont {Sugiyama},\ and\ \citenamefont {Freese}}]{savage2005age}%
  \BibitemOpen
  \bibfield  {author} {\bibinfo {author} {\bibfnamefont {C.}~\bibnamefont {Savage}}, \bibinfo {author} {\bibfnamefont {N.}~\bibnamefont {Sugiyama}},\ and\ \bibinfo {author} {\bibfnamefont {K.}~\bibnamefont {Freese}},\ }\href@noop {} {\bibfield  {journal} {\bibinfo  {journal} {Journal of Cosmology and Astroparticle Physics}\ }\textbf {\bibinfo {volume} {2005}}\bibinfo  {number} { (10)},\ \bibinfo {pages} {007}}\BibitemShut {NoStop}%
\bibitem [{\citenamefont {Doran}\ \emph {et~al.}(2007)\citenamefont {Doran}, \citenamefont {Stern},\ and\ \citenamefont {Thommes}}]{doran2007baryon}%
  \BibitemOpen
\bibfield  {number} {  }\bibfield  {author} {\bibinfo {author} {\bibfnamefont {M.}~\bibnamefont {Doran}}, \bibinfo {author} {\bibfnamefont {S.}~\bibnamefont {Stern}},\ and\ \bibinfo {author} {\bibfnamefont {E.}~\bibnamefont {Thommes}},\ }\href@noop {} {\bibfield  {journal} {\bibinfo  {journal} {Journal of Cosmology and Astroparticle Physics}\ }\textbf {\bibinfo {volume} {2007}}\bibinfo  {number} { (04)},\ \bibinfo {pages} {015}}\BibitemShut {NoStop}%
\bibitem [{\citenamefont {Komatsu}\ \emph {et~al.}(2009)\citenamefont {Komatsu}, \citenamefont {Dunkley}, \citenamefont {Nolta}, \citenamefont {Bennett}, \citenamefont {Gold}, \citenamefont {Hinshaw}, \citenamefont {Jarosik}, \citenamefont {Larson}, \citenamefont {Limon}, \citenamefont {Page} \emph {et~al.}}]{komatsu2009five}%
  \BibitemOpen
\bibfield  {number} {  }\bibfield  {author} {\bibinfo {author} {\bibfnamefont {E.}~\bibnamefont {Komatsu}}, \bibinfo {author} {\bibfnamefont {J.}~\bibnamefont {Dunkley}}, \bibinfo {author} {\bibfnamefont {M.}~\bibnamefont {Nolta}}, \bibinfo {author} {\bibfnamefont {C.}~\bibnamefont {Bennett}}, \bibinfo {author} {\bibfnamefont {B.}~\bibnamefont {Gold}}, \bibinfo {author} {\bibfnamefont {G.}~\bibnamefont {Hinshaw}}, \bibinfo {author} {\bibfnamefont {N.}~\bibnamefont {Jarosik}}, \bibinfo {author} {\bibfnamefont {D.}~\bibnamefont {Larson}}, \bibinfo {author} {\bibfnamefont {M.}~\bibnamefont {Limon}}, \bibinfo {author} {\bibfnamefont {L.}~\bibnamefont {Page}}, \emph {et~al.},\ }\href@noop {} {\bibfield  {journal} {\bibinfo  {journal} {The Astrophysical Journal Supplement Series}\ }\textbf {\bibinfo {volume} {180}},\ \bibinfo {pages} {330} (\bibinfo {year} {2009})}\BibitemShut {NoStop}%
\bibitem [{\citenamefont {Abraham}\ \emph {et~al.}(2004)\citenamefont {Abraham}, \citenamefont {Glazebrook}, \citenamefont {McCarthy}, \citenamefont {Crampton}, \citenamefont {Murowinski}, \citenamefont {J{\o}rgensen}, \citenamefont {Roth}, \citenamefont {Hook}, \citenamefont {Savaglio}, \citenamefont {Chen} \emph {et~al.}}]{abraham2004gemini}%
  \BibitemOpen
  \bibfield  {author} {\bibinfo {author} {\bibfnamefont {R.~G.}\ \bibnamefont {Abraham}}, \bibinfo {author} {\bibfnamefont {K.}~\bibnamefont {Glazebrook}}, \bibinfo {author} {\bibfnamefont {P.~J.}\ \bibnamefont {McCarthy}}, \bibinfo {author} {\bibfnamefont {D.}~\bibnamefont {Crampton}}, \bibinfo {author} {\bibfnamefont {R.}~\bibnamefont {Murowinski}}, \bibinfo {author} {\bibfnamefont {I.}~\bibnamefont {J{\o}rgensen}}, \bibinfo {author} {\bibfnamefont {K.}~\bibnamefont {Roth}}, \bibinfo {author} {\bibfnamefont {I.~M.}\ \bibnamefont {Hook}}, \bibinfo {author} {\bibfnamefont {S.}~\bibnamefont {Savaglio}}, \bibinfo {author} {\bibfnamefont {H.-W.}\ \bibnamefont {Chen}}, \emph {et~al.},\ }\href@noop {} {\bibfield  {journal} {\bibinfo  {journal} {The Astronomical Journal}\ }\textbf {\bibinfo {volume} {127}},\ \bibinfo {pages} {2455} (\bibinfo {year} {2004})}\BibitemShut {NoStop}%
\bibitem [{\citenamefont {Percival}\ \emph {et~al.}(2010)\citenamefont {Percival}, \citenamefont {Reid}, \citenamefont {Eisenstein}, \citenamefont {Bahcall}, \citenamefont {Budavari}, \citenamefont {Frieman}, \citenamefont {Fukugita}, \citenamefont {Gunn}, \citenamefont {Ivezi{\'c}}, \citenamefont {Knapp} \emph {et~al.}}]{percival2010baryon}%
  \BibitemOpen
  \bibfield  {author} {\bibinfo {author} {\bibfnamefont {W.~J.}\ \bibnamefont {Percival}}, \bibinfo {author} {\bibfnamefont {B.~A.}\ \bibnamefont {Reid}}, \bibinfo {author} {\bibfnamefont {D.~J.}\ \bibnamefont {Eisenstein}}, \bibinfo {author} {\bibfnamefont {N.~A.}\ \bibnamefont {Bahcall}}, \bibinfo {author} {\bibfnamefont {T.}~\bibnamefont {Budavari}}, \bibinfo {author} {\bibfnamefont {J.~A.}\ \bibnamefont {Frieman}}, \bibinfo {author} {\bibfnamefont {M.}~\bibnamefont {Fukugita}}, \bibinfo {author} {\bibfnamefont {J.~E.}\ \bibnamefont {Gunn}}, \bibinfo {author} {\bibfnamefont {{\v{Z}}.}~\bibnamefont {Ivezi{\'c}}}, \bibinfo {author} {\bibfnamefont {G.~R.}\ \bibnamefont {Knapp}}, \emph {et~al.},\ }\href@noop {} {\bibfield  {journal} {\bibinfo  {journal} {Monthly Notices of the Royal Astronomical Society}\ }\textbf {\bibinfo {volume} {401}},\ \bibinfo {pages} {2148} (\bibinfo {year} {2010})}\BibitemShut {NoStop}%
\bibitem [{\citenamefont {Planck~Collaboration}\ \emph {et~al.}(2020)\citenamefont {Planck~Collaboration}, \citenamefont {Akrami}, \citenamefont {Ashdown}, \citenamefont {Aumont}, \citenamefont {Baccigalupi}, \citenamefont {Ballardini}, \citenamefont {Banday}, \citenamefont {Barreiro}, \citenamefont {Bartolo}, \citenamefont {Basak} \emph {et~al.}}]{aghanim2020planck}%
  \BibitemOpen
  \bibfield  {author} {\bibinfo {author} {\bibfnamefont {N.}~\bibnamefont {Planck~Collaboration}, \bibfnamefont {Aghanim}}, \bibinfo {author} {\bibfnamefont {Y.}~\bibnamefont {Akrami}}, \bibinfo {author} {\bibfnamefont {M.}~\bibnamefont {Ashdown}}, \bibinfo {author} {\bibfnamefont {J.}~\bibnamefont {Aumont}}, \bibinfo {author} {\bibfnamefont {C.}~\bibnamefont {Baccigalupi}}, \bibinfo {author} {\bibfnamefont {M.}~\bibnamefont {Ballardini}}, \bibinfo {author} {\bibfnamefont {A.}~\bibnamefont {Banday}}, \bibinfo {author} {\bibfnamefont {R.}~\bibnamefont {Barreiro}}, \bibinfo {author} {\bibfnamefont {N.}~\bibnamefont {Bartolo}}, \bibinfo {author} {\bibfnamefont {S.}~\bibnamefont {Basak}}, \emph {et~al.},\ }\href@noop {} {\bibfield  {journal} {\bibinfo  {journal} {Astronomy \& Astrophysics}\ }\textbf {\bibinfo {volume} {641}},\ \bibinfo {pages} {A6} (\bibinfo {year} {2020})}\BibitemShut {NoStop}%
\bibitem [{\citenamefont {Kosowsky}(2003)}]{kosowsky2003atacama}%
  \BibitemOpen
  \bibfield  {author} {\bibinfo {author} {\bibfnamefont {A.}~\bibnamefont {Kosowsky}},\ }\href@noop {} {\bibfield  {journal} {\bibinfo  {journal} {New Astronomy Reviews}\ }\textbf {\bibinfo {volume} {47}},\ \bibinfo {pages} {939} (\bibinfo {year} {2003})}\BibitemShut {NoStop}%
\bibitem [{\citenamefont {Stobie}\ \emph {et~al.}(2000)\citenamefont {Stobie}, \citenamefont {Meiring},\ and\ \citenamefont {Buckley}}]{stobie2000design}%
  \BibitemOpen
  \bibfield  {author} {\bibinfo {author} {\bibfnamefont {R.}~\bibnamefont {Stobie}}, \bibinfo {author} {\bibfnamefont {J.~G.}\ \bibnamefont {Meiring}},\ and\ \bibinfo {author} {\bibfnamefont {D.~A.}\ \bibnamefont {Buckley}},\ }in\ \href@noop {} {\emph {\bibinfo {booktitle} {Optical Design, Materials, Fabrication, and Maintenance}}},\ Vol.\ \bibinfo {volume} {4003}\ (\bibinfo {organization} {International Society for Optics and Photonics},\ \bibinfo {year} {2000})\ pp.\ \bibinfo {pages} {355--362}\BibitemShut {NoStop}%
\bibitem [{\citenamefont {Foreman-Mackey}\ \emph {et~al.}(2013)\citenamefont {Foreman-Mackey}, \citenamefont {Hogg}, \citenamefont {Lang},\ and\ \citenamefont {Goodman}}]{emcee2013}%
  \BibitemOpen
  \bibfield  {author} {\bibinfo {author} {\bibfnamefont {D.}~\bibnamefont {Foreman-Mackey}}, \bibinfo {author} {\bibfnamefont {D.~W.}\ \bibnamefont {Hogg}}, \bibinfo {author} {\bibfnamefont {D.}~\bibnamefont {Lang}},\ and\ \bibinfo {author} {\bibfnamefont {J.}~\bibnamefont {Goodman}},\ }\href {https://doi.org/10.1086/670067} {\bibfield  {journal} {\bibinfo  {journal} {Publications of the Astronomical Society of the Pacific}\ }\textbf {\bibinfo {volume} {125}},\ \bibinfo {pages} {306–312} (\bibinfo {year} {2013})}\BibitemShut {NoStop}%
\bibitem [{\citenamefont {{Hinton}}(2016)}]{Hinton2016}%
  \BibitemOpen
  \bibfield  {author} {\bibinfo {author} {\bibfnamefont {S.~R.}\ \bibnamefont {{Hinton}}},\ }\href {https://doi.org/10.21105/joss.00045} {\bibfield  {journal} {\bibinfo  {journal} {The Journal of Open Source Software}\ }\textbf {\bibinfo {volume} {1}},\ \bibinfo {eid} {00045} (\bibinfo {year} {2016})}\BibitemShut {NoStop}%
\bibitem [{\citenamefont {Valcin}\ \emph {et~al.}(2025)\citenamefont {Valcin}, \citenamefont {Jimenez}, \citenamefont {Seljak},\ and\ \citenamefont {Verde}}]{valcin2025ageuniverseglobularclusters}%
  \BibitemOpen
  \bibfield  {author} {\bibinfo {author} {\bibfnamefont {D.}~\bibnamefont {Valcin}}, \bibinfo {author} {\bibfnamefont {R.}~\bibnamefont {Jimenez}}, \bibinfo {author} {\bibfnamefont {U.}~\bibnamefont {Seljak}},\ and\ \bibinfo {author} {\bibfnamefont {L.}~\bibnamefont {Verde}},\ }\href {https://arxiv.org/abs/2503.19481} {\bibinfo {title} {The age of the universe with globular clusters iii: Gaia distances and hierarchical modeling}} (\bibinfo {year} {2025}),\ \Eprint {https://arxiv.org/abs/2503.19481} {arXiv:2503.19481 [astro-ph.CO]} \BibitemShut {NoStop}%
\end{thebibliography}%

\end{document}